\documentclass{aa}
\bibpunct{(}{)}{;}{a}{}{,} 
\usepackage{graphicx}
\usepackage{subfig}
\usepackage{comment}
\usepackage{xcolor}
\usepackage{ulem}

\newcommand{\teff}{$T_{\rm eff}$}
\newcommand{\logg}{$\log g$}
\newcommand{\vsini}{$v\sin i$}

\title{PENELLOPE V. The magnetospheric structure and the accretion variability of the classical T~Tauri~star HM Lup\thanks{Based on observations collected at the European Southern Observatory under ESO programmes 089.C-0143(A), 106.20Z8.003, and 106.20Z8.004.}}
\author{A. Armeni\inst{\ref{inst1}} \and B. Stelzer\inst{\ref{inst1},\ref{inst2}} \and R.~A.~B. Claes\inst{\ref{inst3}} \and C.~F. Manara\inst{\ref{inst3}} \and A. Frasca\inst{\ref{inst4}} \and J.~M. Alcalá\inst{\ref{inst5}} \and F.~M. Walter\inst{\ref{inst6}} \and \'A.~K\'osp\'al\inst{\ref{inst7},\ref{inst8},\ref{inst9},\ref{inst10}} \and J. Campbell-White\inst{\ref{inst3}} \and M.~Gangi\inst{\ref{inst11},\ref{inst12}} \and K. Mauco\inst{\ref{inst3}} \and L. Tychoniec\inst{\ref{inst3}}}

\institute{
    Institut für Astronomie und Astrophysik, Eberhard-Karls Universität Tübingen, Sand 1, 72076 Tübingen, Germany \email{armeni@astro.uni-tuebingen.de} \label{inst1}
    \and
    INAF - Osservatorio Astronomico di Palermo, Piazza del Parlamento 1, 90134 Palermo, Italy \label{inst2}
    \and
    European Southern Observatory, Karl-Schwarzschild-Strasse 2, 85748 Garching bei München, Germany
    \label{inst3}
    \and
    INAF – Osservatorio Astrofisico di Catania, via S. Sofia 78, 95123 Catania, Italy
    \label{inst4}
    \and
    INAF – Osservatorio Astronomico di Capodimonte, via Moiariello 16, 80131 Napoli, Italy
    \label{inst5}
    \and
    Department of Physics \& Astronomy, Stony Brook University,
    Stony Brook NY 11794-3800, USA
    \label{inst6}
    \and
    Konkoly Observatory, Research Centre for Astronomy and Earth Sciences, E\"otv\"os Lor\'and Research Network (ELKH), Konkoly-Thege Mikl\'os \'ut 15-17, H-1121 Budapest, Hungary
    \label{inst7}
    \and
    CSFK, MTA Centre of Excellence, Konkoly Thege Miklós út 15-17, H-1121, Budapest, Hungary
    \label{inst8}
    \and
    ELTE E\"otv\"os Lor\'and University, Institute of Physics, P\'azm\'any P\'eter s\'et\'any 1/A, H-1117 Budapest, Hungary
    \label{inst9}
    \and
    Max-Planck-Institut f\"ur Astronomie, K\"onigstuhl 17, 69117 Heidelberg, Germany
    \label{inst10}
    \and
    INAF – Osservatorio Astronomico di Roma, via Frascati 33, 00078 Monte Porzio Catone, Italy
    \label{inst11}
    \and
    ASI, Italian Space Agency, Via del Politecnico snc, 00133 Rome, Italy
    \label{inst12}
}

\date{Received: 30 May 2023 / Accepted: 15 September 2023}

\abstract {HM Lup is a young M-type star that accretes material from a circumstellar disk through a magnetosphere. Our aim is to study the inner disk structure of HM Lup and to characterize its variability. We used spectroscopic data from HST/STIS, X-Shooter, and ESPRESSO taken in the framework of the ULLYSES and PENELLOPE programs, together with photometric data from TESS and AAVSO. The 2021 TESS light curve shows variability typical for young stellar objects of the "accretion burster" type. The spectra cover the temporal evolution of the main burst in the 2021 TESS light curve.
We compared the strength and morphology of emission lines from different species and ionization stages. We determined the mass accretion rate from selected emission lines and from the UV continuum excess emission at different epochs, and we examined its relation to the photometric light curves.
The emission lines in the optical spectrum of HM\,Lup delineate a temperature stratification along the accretion flow. While the wings of the \ion{H}{i} and \ion{He}{i} lines originate near the star, the lines of species such as \ion{Na}{i}, \ion{Mg}{i}, \ion{Ca}{i}, \ion{Ca}{ii}, \ion{Fe}{i}, and \ion{Fe}{ii} are formed in an outer and colder region.
The shape and periodicity of the 2019 and 2021 TESS light curves, when qualitatively compared to predictions from magnetohydrodynamic models, suggest that HM\,Lup was in a regime of unstable ordered accretion during the 2021 TESS observation due to an increase in the accretion rate.
Although HM\,Lup is not an extreme accretor, it shows enhanced emission in the metallic species during this high accretion state that is produced by a density enhancement in the outer part of the accretion flow.} 


\begin{document}
	\titlerunning{The magnetospheric structure and accretion variability of HM Lup}
	\authorrunning{A. Armeni et al.}
	\maketitle
	\section{Introduction}
	Classical T Tauri stars (CTTSs) are young ($\sim 1-10$ Myr), low-mass ($< 2 \, M_{\odot}$) objects surrounded by a circumstellar disk \citep{Hartmann+2016}. Their strong magnetic fields  truncate the disk at a few stellar radii (typically 5~$R_{\star}$, \citealt{Hartmann+1998}). The current paradigm for the interaction between the disk and the star is the magnetospheric accretion \citep{Bouvier+2007}, in which the material free-falls onto the star following the magnetic field lines. 
	The rich emission line spectrum typical of CTTSs \citep{Joy1945, Herbig1962} can be explained in the framework of this model \citep{Hartmann+1994, Muzerolle+1998}, as well as the continuum excess flux, which results from the accretion shock at the stellar surface \citep{CalvetGullbring1998}.
	
	Young stellar objects (YSOs) are known to be variable, both photometrically and spectroscopically \citep{Joy1945, Herbst+1994, Hartmann+2016, Fischer+2022}. Many different processes can contribute to this variability, such as variable accretion rate, rotational modulation due to stellar spots, circumstellar extinction, and flares \citep{Cody+2014, Cody+2022}. 
	The accretion process is often accompanied by outflows \citep{Hartmann+2016, Bally2016}, either in the form of disk winds or jets \citep[e.g.,][and references therein]{Romanova+2009, Ferreira2013}. 
	
	Since the spectro-photometric variability of YSOs shows up in a broad range of wavelengths, from the X-rays to the infrared \citep{AppenzellerMundt1989}, it is essential to study these objects using a multiwavelength approach, by means of simultaneous observations in different spectral regions. 
	Many works have shown the capabilities of simultaneous spectro-photometry in determining stellar and accretion parameters, unveiling the inner disk structure of CTTSs and studying accretion variability on different timescales \citep[e.g.,][]{Bouvier+2007b, Alencar+2018, Zsidi+2022a, Zsidi+2022b, Fiorellino+2022}.
	This requires coordinated monitoring campaigns of a range of instruments, which are notoriously difficult to achieve. 
	The \textit{Hubble UV Legacy Library of Young Stars as Essential Standards}, \citep[ULLYSES,][]{RomanDuval+2020, Espaillat+2022} now offers such a possibility. The aim of this program is to obtain low and medium resolution spectra of YSOs covering the wavelength range from the far-UV ($\sim 150$~nm) to the infrared ($\sim 1000$~nm) with the Hubble Space Telescope (HST).
	Together with the accompanying optical program PENELLOPE \citep{Manara+2021} at ESO Very Large Telescope (VLT), it provides an unprecedented spectroscopic dataset to study the accretion variability of YSOs. 
	
	PENELLOPE typically provides three high resolution spectra either with the {\it Echelle SPectrograph for Rocky Exoplanets and Stable Spectroscopic Observations} \cite[ESPRESSO,][]{Pepe+2021} or the {\it Ultraviolet and Visual Echelle Spectrograph} \citep[UVES,][]{Dekker+2000} and one medium resolution X-Shooter \citep{Vernet+2011} spectrum per target taken close in time to the ULLYSES observations. The high resolution spectra are needed to study variability in the emission line profiles, while the X-Shooter spectrum is used to constrain the stellar and accretion parameters.
	As much as possible, the ULLYSES and PENELLOPE observations are scheduled during times where the targets are observed with the Transiting Exoplanet Survey Satellite \citep[TESS,][]{Ricker+2014}. 
	TESS produces short-cadence, about one-month-long light curves for many YSOs with a spectral response that covers the red/infrared wavelength range ($\sim 0.6 - 1.1 \, \rm{\mu m}$).
	Since most of the excess flux due to accretion is emitted in the near-UV to optical spectral region, at wavelengths shorter than the TESS filter band \citep{CalvetGullbring1998}, it is also important to obtain simultaneous multiband photometry \citep{Robinson+2022}. 
	
	In this work we take advantage of the wealth of data provided by the aforementioned programs to study in detail a single target, chosen because of the simultaneous coverage with all the relevant programs.
	The target of this study is HM Lup (Sz 72), a CTTS with spectral type M2 and a mass of $0.37 M_{\rm{\odot}}$ \citep{Alcala+2014, Manara+2022} located in the Lupus cloud at a distance of $156$~pc \citep{GaiaDR3}. It hosts a disk with a dust mass of $4.09 ~ M_{\oplus}$ and a radius of $10.97$~au \citep{Manara+2022}, inclined by $53 \pm 19$~deg relative to the line of sight \citep{Ansdell+2016}. 
	
	The goal of this paper is twofold. First, we aim to study the temperature stratification of the magnetosphere of HM\,Lup using a range of emission lines. Secondly, we aim to investigate the spectrophotometric variability of the system.
	
	The paper is structured as follows. In Sect.~\ref{obs} we describe the observations.
	We report the basic properties of the star in Sect.~\ref{stellar_pars}. We present the results based on the optical spectrum of the system in Sect.~\ref{opt_spec} and the results on the spectrophotometric variability in Sect.~\ref{spectrophotometry}, while we discuss them in Sect.~\ref{discussion}.
	
	\section{Observations}
	\label{obs}
	For our study, we selected HM\,Lup as one of the PENELLOPE targets having spectra taken simultaneously with a TESS observation, in this case in sector 38 (2021). 
	HM\,Lup was also observed with TESS in Sector~12 (2019). 
	
	\subsection{Simultaneous data}
	\begin{figure*}
		\centering
		\includegraphics[width=\linewidth]{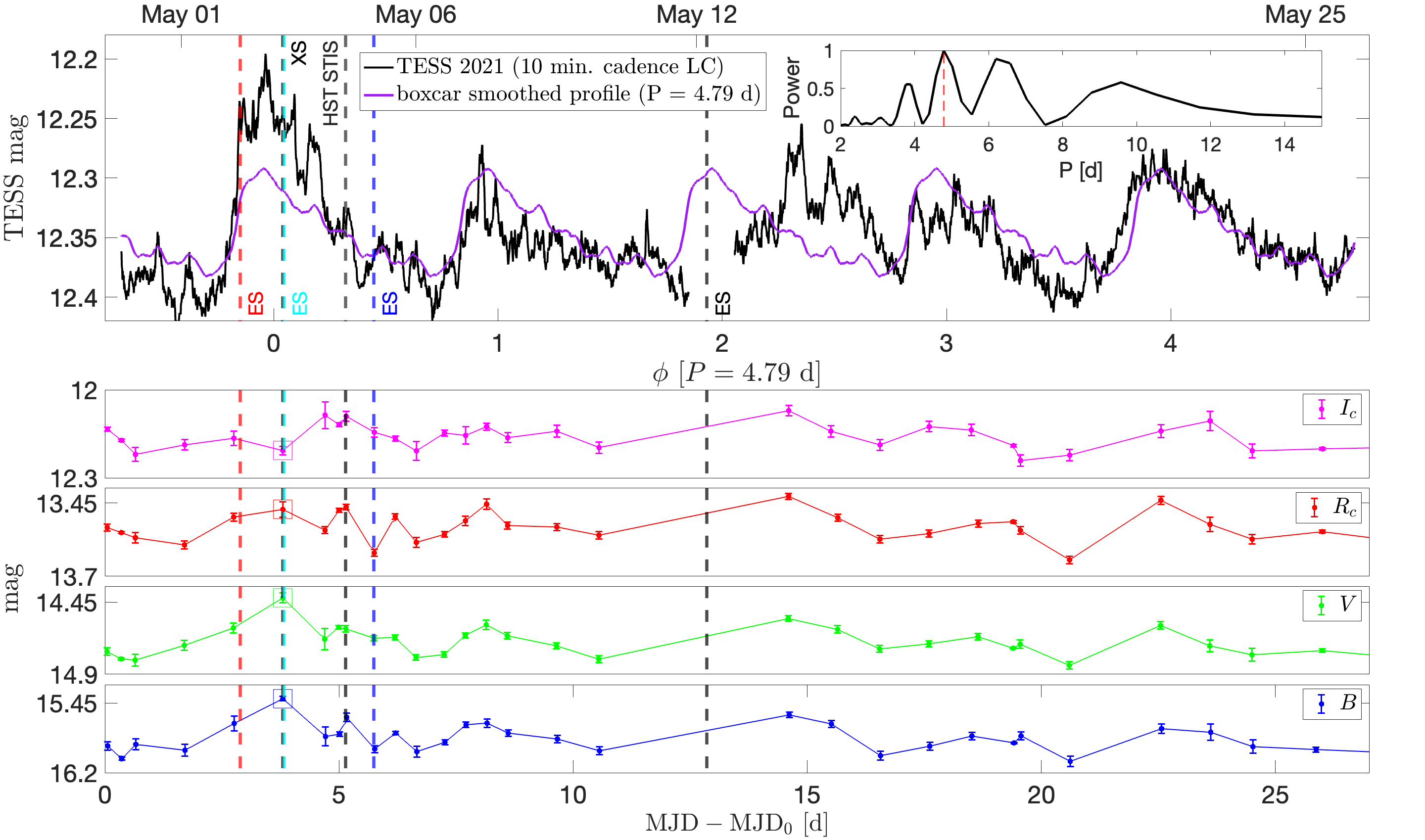}
		\caption{Light curves and timing of simultaneous spectroscopy for HM Lup. The topmost panel shows the 2021 10 minute cadence TESS light curve (Sector\,38). The inset displays the Lomb-Scargle Periodogram for the TESS light curve. Overlaid in purple is a boxcar smoothed version of the light curve produced by phase-folding the data with the period of the highest peak in the Lomb-Scargle periodogram, $4.79$~d. 
		The phase $\phi$ was computed with this period and $\phi = 0$ at the maximum of the TESS light curve, MJD~$59336.97$.
		The vertical dashed lines mark the epochs of the simultaneous spectroscopic observations. Here ES $=$ ESPRESSO and XS $=$ X-Shooter. 
		The other panels show the part of the AAVSO $BVR_{\rm{c}}I_{\rm{c}}$ photometry that is simultaneous with TESS, with a linear interpolation as a guideline. The large open squares mark the synthetic photometry obtained from the X-Shooter spectrum for the four filters. We defined $\rm{MJD_0} = 59333.363$ as the beginning of TESS Sector\,38 observation.
		}
		\label{fig:simultaneous_spectrophot}
	\end{figure*}
	
	We downloaded the TESS light curves of sectors 12 and 38 from the MAST archive\footnote{\url{https://archive.stsci.edu/}}. 
	The full frame images (FFI) were reduced by the TESS Science Processing Operations Center (SPOC), producing a 10 minute cadence light curve. 
	We define the beginning of TESS Sector 38 observations, $\rm{MJD_0}\equiv 59333.363$, as the reference time for all the simultaneous spectroscopic and photometric data.
    Figure~\ref{fig:simultaneous_spectrophot} shows the TESS light curve in which the Pre-search Data Conditioning Simple Aperture Photometry (PDCSAP) flux ($F$) was converted into TESS magnitudes ($T$) using the relation $T = -2.5\cdot\log_{10}F + ZP$ where $ZP = 20.44$ is the TESS Zero Point magnitude \citep{Fausnaugh+2021, Vanderspek+2018}.

	We downloaded $BVR_{\rm c}I_{\rm c}$ photometry available for HM\,Lup from the American Association of Variable Star Observers (AAVSO) International Database\footnote{\url{https://www.aavso.org/aavso-international-database-aid}}. Most of the data were secured from the beginning of TESS observations and span $\sim 140$ days. 
	Data from different observers were sometimes obtained during the same day at small temporal distance ($\Delta t \sim 15 \, \rm{min}$). To have a better view of the variability, we binned the photometry to $0.15$\,d. 

	HM Lup was selected as an ULLYSES target and observed on 4 May 2021 with the Space Telescope Imaging Spectrograph (STIS) camera between $1650$ and $10200$~{\AA} with a resolving power of $R\sim1500$.
	Contemporaneous to the HST data, medium-high resolution optical spectroscopy was obtained in the framework of PENELLOPE \citep{Manara+2021}. The journal of the spectroscopic observations is reported in Table~\ref{tab:journal_obs}, while Fig.~\ref{fig:simultaneous_spectrophot} shows the temporal position of the spectra relative to the 2021 photometric data.
	High resolution spectra were obtained with ESPRESSO in Pr.~Id. 106.20Z8.003 (PI Manara). Since in the first observation the conditions did not fulfill the requirements, namely the seeing was $>1.55^{\prime\prime}$, the observation was repeated the day after. However, with its signal-to-noise ratio (S/N) of $12$, the first spectrum can still be used. The ESPRESSO spectra cover a wavelength range between $3800$ and $7880$~{\AA} with a resolution of $140000$. 
	These spectra were flux-calibrated as explained in Appendix~\ref{ESPRESSO_flux_calib}. 
	The X-Shooter spectrum taken from Pr.~Id. 106.20Z8.004 (PI Manara) is divided into three arms, UVB ($3000 - 5600$~{~\AA}, $R \sim 5400$), VIS ($5600 - 10200${~\AA}, $R \sim 18400$), and NIR ($10200 - 24800${~\AA}, $R \sim 11600$). In order to achieve this spectral resolution, the observation was carried out with slit widths of $1^{\prime\prime}$, $0.4^{\prime\prime}$, $0.4^{\prime\prime}$ in the UVB, VIS and NIR arms, respectively. 
	The absolute flux calibration correcting for slit losses was achieved using a short exposure taken with a $5^{\prime\prime}$ slit, as explained by \citet{Manara+2021}.
	Telluric correction was performed on the ESPRESSO and X-Shooter spectra using the molecfit tool \citep{Smette+2015}.
	All the optical spectra were reduced by the PENELLOPE team and made publicly available on Zenodo\footnote{\url{https://zenodo.org/communities/odysseus/}}.

    \subsection{Nonsimultaneous data}
    \begin{figure*}
		\centering
		\includegraphics[width=\linewidth]{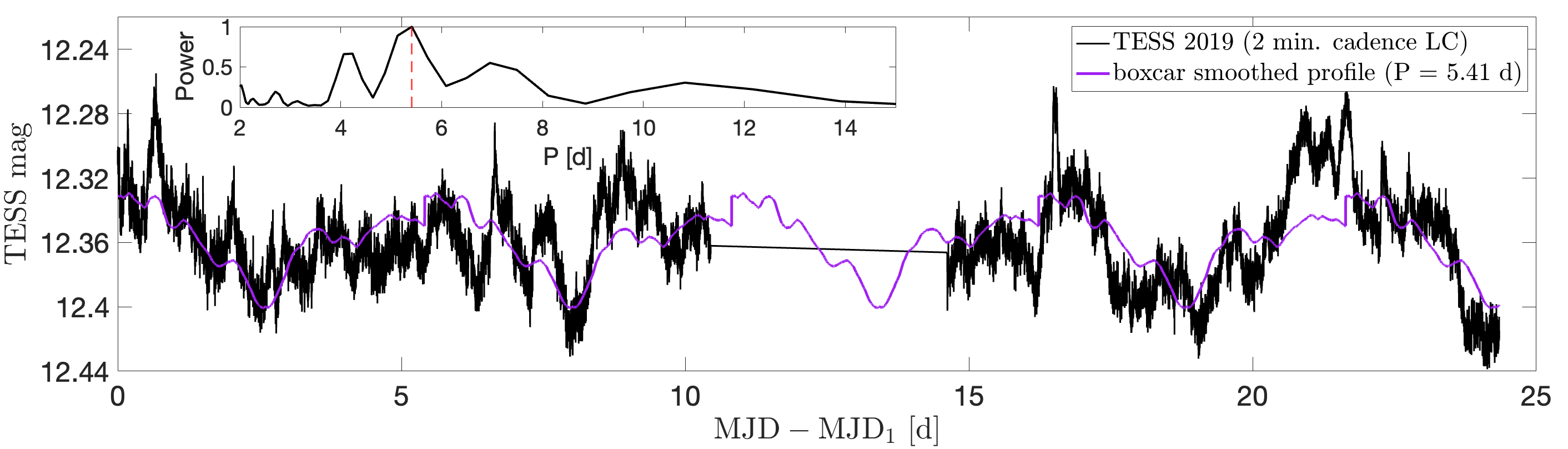}
		\caption{TESS 2019 two-minute cadence light curve (Sector\,12) of HM~Lup. The inset shows the Lomb-Scargle periodogram. Overlaid in purple is a boxcar smoothed version of the light curve produced by phase-folding the data with the period of the highest peak in the Lomb-Scargle periodogram, $5.41$~d. We defined $\rm{MJD_1} \equiv 58628.050$ as the beginning of the TESS Sector\,12 observation. 
		}
		\label{fig:TESS_2019}
	\end{figure*}
    In addition to the simultaneous spectro-photometric data, we have also included observations of HM\,Lup taken at other epochs. In particular, an X-Shooter spectrum obtained on 18 April 2012, which was already presented by \citet{Alcala+2014, Alcala+2017} and \citet{Frasca+2017} taken from Pr.~Id. 089.C-0143(A) (PI Alcalá), and the TESS light curve from sector 12, obtained on 25 May 2019, UT 01:11:43 $(\rm MJD_1 \equiv 58628.050)$ and shown in Fig.~\ref{fig:TESS_2019}.
    
    \begin{table}
    	\centering
    	\caption{Journal of the spectroscopic observations. The exposure time values for the X-Shooter spectra are for the UVB, VIS and NIR arms respectively. MJD$_0 = 59333.363$.} 
    	\begin{tabular}{ccc}
    	    \hline
    	    Instrument & $\rm{MJD-MJD_0}$~[d] & $t_{\rm{exp}}$~[s] \\
    	    \hline
    	    X-Shooter & $-3298.17$ & $300-250-100$ \\
    	    ESPRESSO & $2.89$ & $1650$ \\
    	    X-Shooter & $3.80$ & $470-380-100$ \\
    	    ESPRESSO & $3.83$ & $1650$ \\
    	    HST/STIS & $5.14$ & $1230$ \\
    	    ESPRESSO & $5.75$ & $1650$ \\
    	    ESPRESSO & $12.86$ & $1650$ \\
    		\hline
    	\end{tabular}
    	\label{tab:journal_obs} 
    \end{table}
    
	
	\section{Stellar parameters}
	\label{stellar_pars}
	
	\begin{table}
    	\centering
    	\caption{Stellar parameters of HM\,Lup obtained by fitting the spectra as described in Sect~\ref{stellar_pars}.} 
    	\begin{tabular}{ccc}
    	    \hline
    	    Parameter & Value & Reference \\
    	    \hline
    	    \teff$^{\dagger}$ & $3550 \pm 70$~K & \citet{Frasca+2017} \\
    	    \logg$^{\dagger}$ & $4.18 \pm 0.28$~dex & \citet{Frasca+2017}  \\
    	    $L_{\star}^{\dagger}$ & $0.27 \pm 0.13$~$L_{\odot}$ & \citet{Manara+2022} \\
    	    $M_{\star}^{\dagger}$ & $0.37 \pm 0.12$~$M_{\odot}$ & \citet{Manara+2022} \\
    	    $R_{\star}^{\dagger}$ & $1.39 \pm 0.34$~$R_{\odot}$ & this work \\
    	    \vsini$^{\ddagger}$ & $5.8 \pm 0.4~\rm{km~s^{-1}}$ & this work  \\
    	    RV$^{\ddagger}$ & $-2.4 \pm 0.6~\rm{km~s^{-1}}$ & this work \\
    	    $i_{\rm{d}}$ & $53 \pm 19$~deg & \citet{Ansdell+2016} \\ 
    		\hline
    	\end{tabular}
    	\tablefoot{Parameters derived from the 2012 X-Shooter spectrum $(^{\dagger})$ and the 2021 ESPRESSO spectra $(^{\ddagger})$.
    	}
    	\label{tab:stellar_accretion_pars} 
    \end{table}
	
	The properties of the system were determined by \citet{Alcala+2017} and \citet{Frasca+2017} by fitting the 2012 X-Shooter spectrum. 
	While the first paper focused on the spectral type (SpT), the stellar luminosity $L_{\star}$ and the accretion properties, the second work used the ROTFIT routine \citep{Frasca+2015} to derive the atmospheric parameters of the accreting star, namely the effective temperature \teff, the gravity \logg, the projected rotational velocity $v\sin i$, and the systemic radial velocity RV.
	\citet{Manara+2022} updated the values from \citet{Alcala+2017} using the {\it Gaia} DR3 distance and assuming the \citet{HerczegHillenbrand2014} relation between SpT and $T_{\rm eff}$ and non-magnetic evolutionary tracks by \citet{Feiden2016}.
	We applied the same procedure to the 2021 X-Shooter and ESPRESSO spectra. Except for the RV, the parameters are in agreement with the previous result. The best fit RV value for the 2012 spectrum was $6.9 \pm 2.4~\rm{km~s^{-1}}$ while we obtained $\rm{RV} = -2.7 \pm 1.9~\rm{km~s^{-1}}$ from the 2021 X-Shooter spectrum, suggesting a possible companion. 
	We measured the RVs in the ESPRESSO spectra, obtaining $-3.1 \pm 0.7~\rm{km~s^{-1}}$, $-3.6 \pm 0.7~\rm{km~s^{-1}}$, $-2.5 \pm 0.3~\rm{km~s^{-1}}$, and $-2.7 \pm 0.5~\rm{km~s^{-1}}$ for the epochs 1, 2, 3, and 4 respectively. Therefore, no RV variation (within 1$\sigma$) emerges from the spectra during the 2021 PENELLOPE campaign.
	We did not find other signs of binarity in the spectra, such as asymmetric absorption lines. In summary from the RV measurements, we cannot exclude that HM\,Lup is actually a long-period ($P \gtrsim 10$~d) single-lined spectroscopic binary.
	
	Table \ref{tab:stellar_accretion_pars} reports the stellar parameters of the system.
	Regarding $v \sin i$ and RV, we adopted the values obtained from an average of the best fit values for the ESPRESSO spectra, given their higher resolution. 
	We computed the stellar radius $R_{\star}$ inverting the relation $L_{\star} = 4\pi R_{\star}^2 \sigma T_{\rm{eff}}^4$. 
	
\begin{figure}
        \centering
        \includegraphics[width=\linewidth]{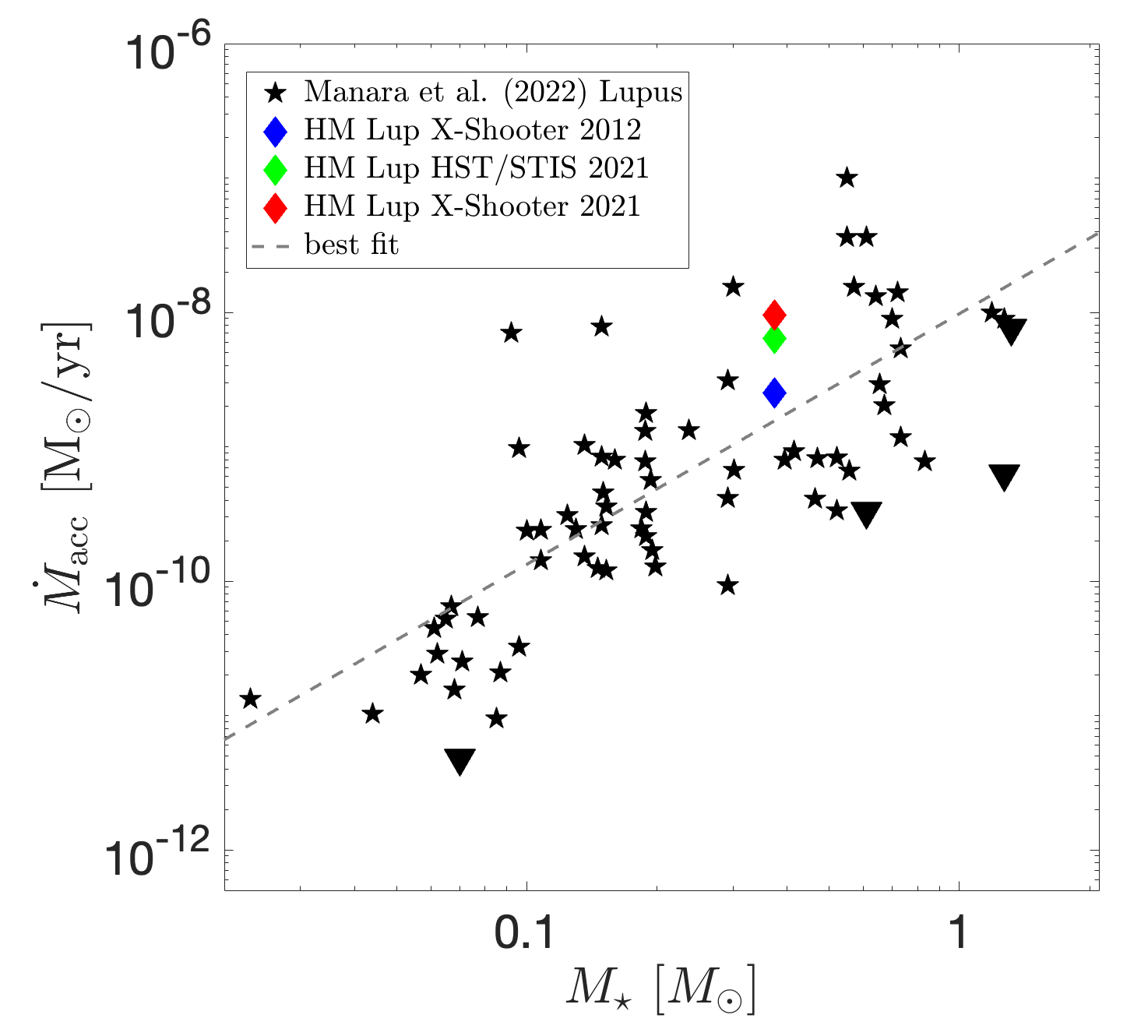}
        \caption{$\dot{M}_{acc}$-$M_{\star}$ diagram for X-Shooter targets in Lupus, using the values from \citet{Manara+2022}. The black triangles indicate upper limits on the measured accretion rate. The blue, green and red diamonds mark the accretion rate of HM\,Lup in 2012 and 2021 (Sect.~\ref{slab_model}). The dashed gray line is a linear best fit to the data.}
        \label{fig:Mstar_Macc}
    \end{figure}

	In Fig.~\ref{fig:Mstar_Macc} we show HM\,Lup in the $\dot{M}_{acc}$-$M_{\star}$ diagram together with the whole X-Shooter Lupus sample. Both the 2012 measurement of the accretion rate and our anticipated new values (from Sect.~\ref{slab_model}) are included, showing that HM\,Lup is a strongly accreting CTTS.

    \section{The optical spectrum of the system}
    \label{opt_spec}
\begin{figure*}
        \centering
        \includegraphics[width=\linewidth]{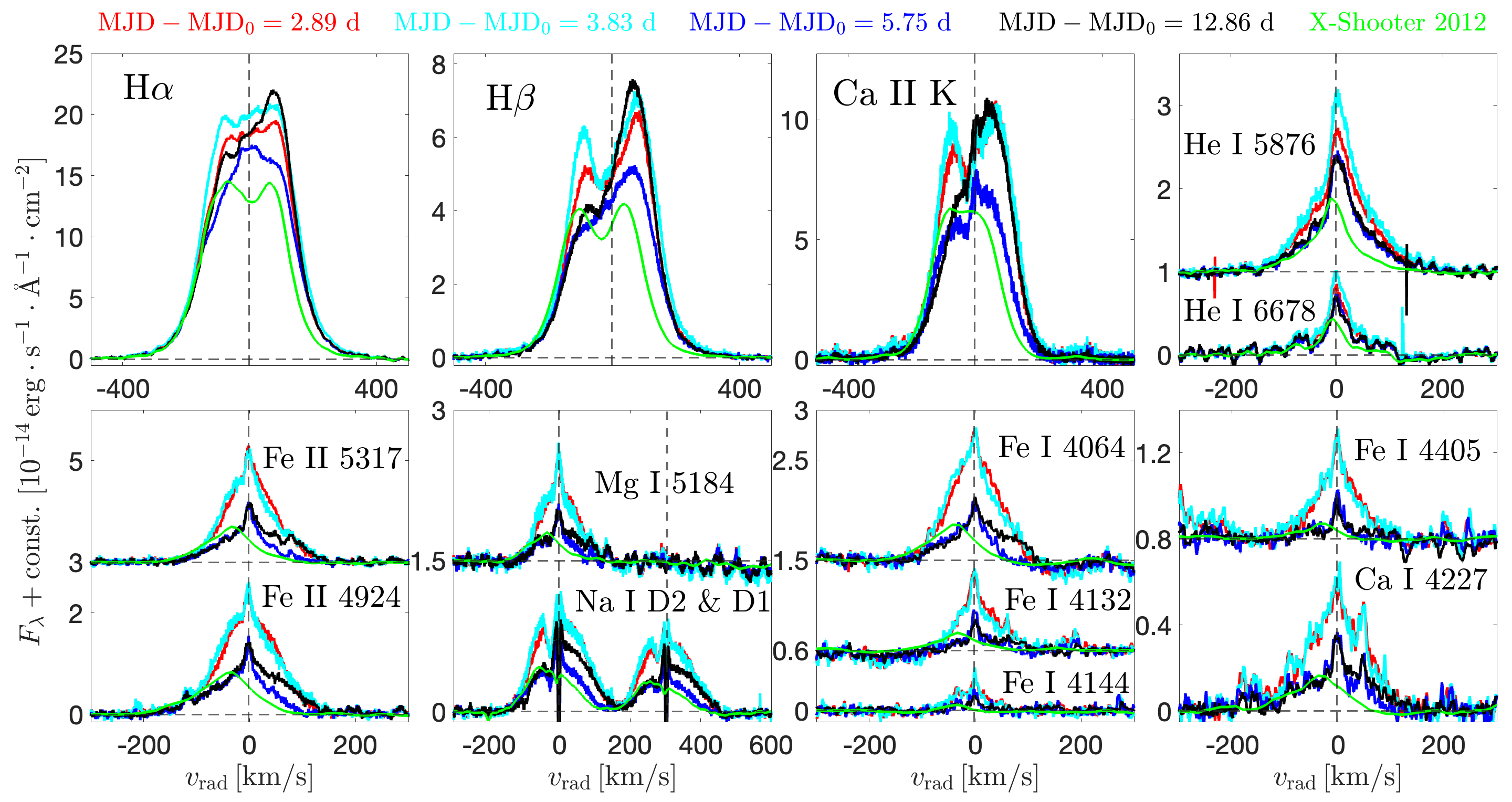}
        \caption{Selection of continuum subtracted emission lines in the ESPRESSO spectra of HM~Lup. The colors for the ESPRESSO spectra are the same as in Fig.~\ref{fig:simultaneous_spectrophot}.
        The lines from the 2012 X-Shooter spectrum are shown in green for comparison. The vertical dashed lines mark the radial velocity of the system, while the horizontal dashed lines highlight the zero-flux level for each set of lines. The \ion{Fe}{i} and \ion{Ca}{i} lines were smoothed with a 7 points boxcar filter.}
        \label{fig:ESPRESSO_LINES}
    \end{figure*}

    The optical spectrum of HM\,Lup is rich in emission lines, the strongest being the Balmer series and the \ion{Ca}{ii} H \& K lines. In addition, permitted emission lines from many other species can be identified, such as \ion{He}{i} and singly and doubly ionized metallic elements, for example, \ion{Na}{i}, \ion{Ca}{i}, \ion{Ti}{i} and \ion{Ti}{ii}, \ion{Fe}{i}, and \ion{Fe}{ii}. We detected more than $150$ emission lines from the iron peak elements. This feature is reminiscent of the outburst spectra of EXors \citep{Sicilia-Aguilar+2012, Sicilia-Aguilar+2017} or the spectra of other strong accretors, such as DR~Tau \citep{Beristain+1998}. We found outflow signatures in the spectra, such as the [\ion{O}{i}]~6300 line in emission and blueshifted absorption in the \ion{He}{i}~10830 line. However, these features will not be analyzed in this work.
	A selection of the observed permitted spectral lines is shown in Fig.~\ref{fig:ESPRESSO_LINES} for the four ESPRESSO spectra. 
	
    The different excitation potentials of the observed transitions and the differences in the emission line profiles indicate the presence of a thermally stratified environment, with multiple regions that contribute to the observed spectrum. In this section we focus on a single spectrum, the epoch 2 of ESPRESSO (cyan in Fig.~\ref{fig:ESPRESSO_LINES}), to illustrate the stratification of the accretion flow. We chose this spectrum because of its high S/N and the strength of the emission lines from low excitation transitions of neutral and singly ionized species.
    We discuss the line variability in Sect.~\ref{line_variability}.
    The atomic parameters for the analyzed emission lines were taken from the NIST Atomic Spectra Database\footnote{\url{https://physics.nist.gov/PhysRefData/ASD/lines_form.html}}.
    
    \subsection{Balmer series, \ion{He}{i}, \ion{Ca}{ii} K}
    \label{HI_CaII_HeI}
    According to the morphological line profile classification by \citet{Reipurth+1996}, the Balmer series, except for H$\alpha$, and the \ion{Ca}{ii} K line have a Type IIB profile, that is, a double-peaked emission profile in which the secondary peak exceeds half the strength of the primary peak. The H$\alpha$ line has instead a flat-topped emission profile. Balmer and \ion{Ca}{ii}~K emission lines have symmetrical wings up to $\pm 400~\rm{km~s^{-1}}$. The \ion{He}{i} lines have less broad wings, up to $\sim \pm 200~\rm{km~s^{-1}}$, and consist of a narrow component (NC) and a broad component (BC). 
    In the rest of the paper, we focus our analysis on the BC emission. 
    
\begin{figure}
		\centering
		\includegraphics[width=\linewidth]{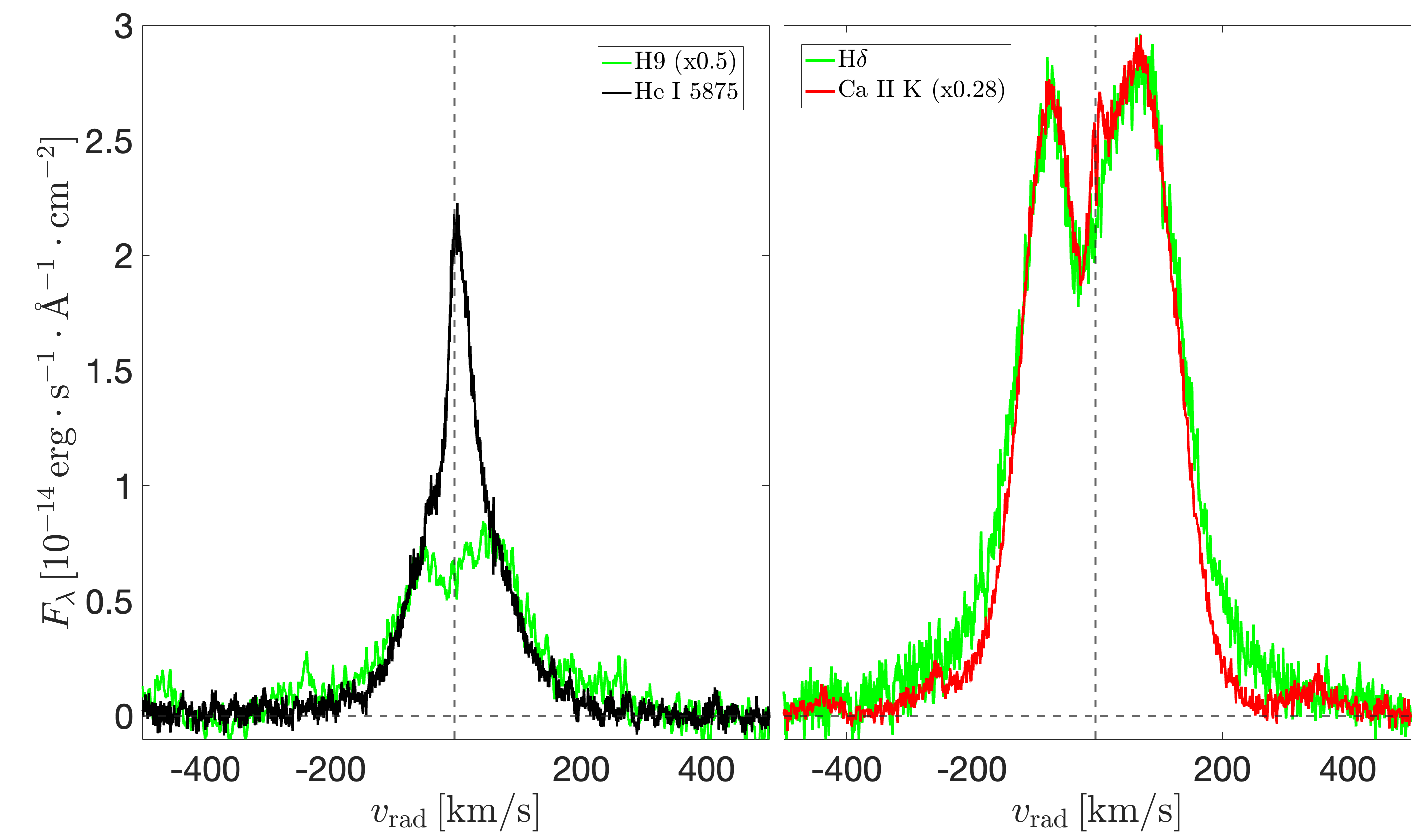}
		\caption{Continuum subtracted profiles of selected magnetospheric lines in the epoch 2 ESPRESSO spectrum. In the left panel, H9 (3835 {\AA}) and \ion{He}{i} 5876. In the right panel, H$\delta$ and \ion{Ca}{ii} K. The vertical dashed line marks the radial velocity of the system. The H9 line was smoothed with a boxcar filter and multiplied by 0.5 to match the \ion{He}{i} 5876 wings. The \ion{Ca}{ii} K line was rescaled to H$\delta$ in a similar way. 
		}
		\label{fig:Balmer_HeI_CaII}
	\end{figure}

    In the magnetospheric accretion scenario, the BC is expected to originate in the infalling material \citep{Beristain+1998, Beristain+2001, Hartmann+2016}.
    The free fall velocity onto a star of mass $M_{\star}$ starting from rest at the disk truncation radius $R_{\rm{T}}$ is
    \begin{equation}
        v_{\rm{ff}} (r) = (2GM_{\star})^{1/2} \left( \frac{1}{r} - \frac{1}{R_{\rm{T}}} \right)^{1/2}.
    \end{equation}
    Assuming $R_{\rm{T}} = 5 ~ R_{\star}$ \citep{Gullbring+1998} and the stellar parameters of HM\,Lup from Table~\ref{tab:stellar_accretion_pars}, we obtain a free fall velocity of $\sim 175~\rm{km~s^{-1}}$ at $r = 2 ~ R_{\star}$, consistent with the observed line wings of the \ion{Ca}{ii} K and \ion{He}{i} 5876 lines. The wings of the lower Balmer lines, especially H$\alpha$ and H$\beta$, exceed these values, possibly due to Stark broadening \citep{Muzerolle+2001, Wilson+2022}. Conversely, the higher lines of the series are less affected by Stark broadening and their wings agree well with the wings of \ion{He}{i} 5876, as shown in the left panel of Fig.~\ref{fig:Balmer_HeI_CaII} for the H9 line. This suggests a common origin for the BC of the \ion{H}{i} and the \ion{He}{i} lines. 
    Given the high excitation potentials of the \ion{He}{i} lines, which have upper states with energies $E_j \gtrsim 20$~eV, a source of ionizing radiation is required \citep{Beristain+2001}. In the magnetospheric accretion flow, these conditions are met in the pre-shock region, that is heated by soft X-rays emitted by the accretion shock \citep{Hartmann+2016}. 
    Singly ionized calcium, with its ionization potential of $11.87$~eV, is almost completely ionized in such conditions. \citet{Azevedo+2006} studied the formation of the \ion{Ca}{ii} infrared triplet lines and showed that the gas departs from LTE conditions. Close to the star, because of the higher dilution factor for the accretion shock radiation, calcium is mostly in the doubly ionized stage (\ion{Ca}{iii}). However, we observe strong \ion{Ca}{ii} emission at all ESPRESSO epochs (Fig.~\ref{fig:ESPRESSO_LINES}). This suggests that the \ion{Ca}{ii} lines are emitted from an outer part of the magnetospheric flow, where calcium is predominantly in the singly ionized stage.
    The right panel of Fig.~\ref{fig:Balmer_HeI_CaII} shows how the wings of the \ion{Ca}{ii} K line are less pronounced than the wings of, for example, H$\delta$, in agreement with this hypothesis.
    
    \subsection{\ion{Fe}{i} fluorescence}
    \label{FeI_fluorescence}
    
    \begin{figure}
		\centering
		\includegraphics[width=\linewidth]{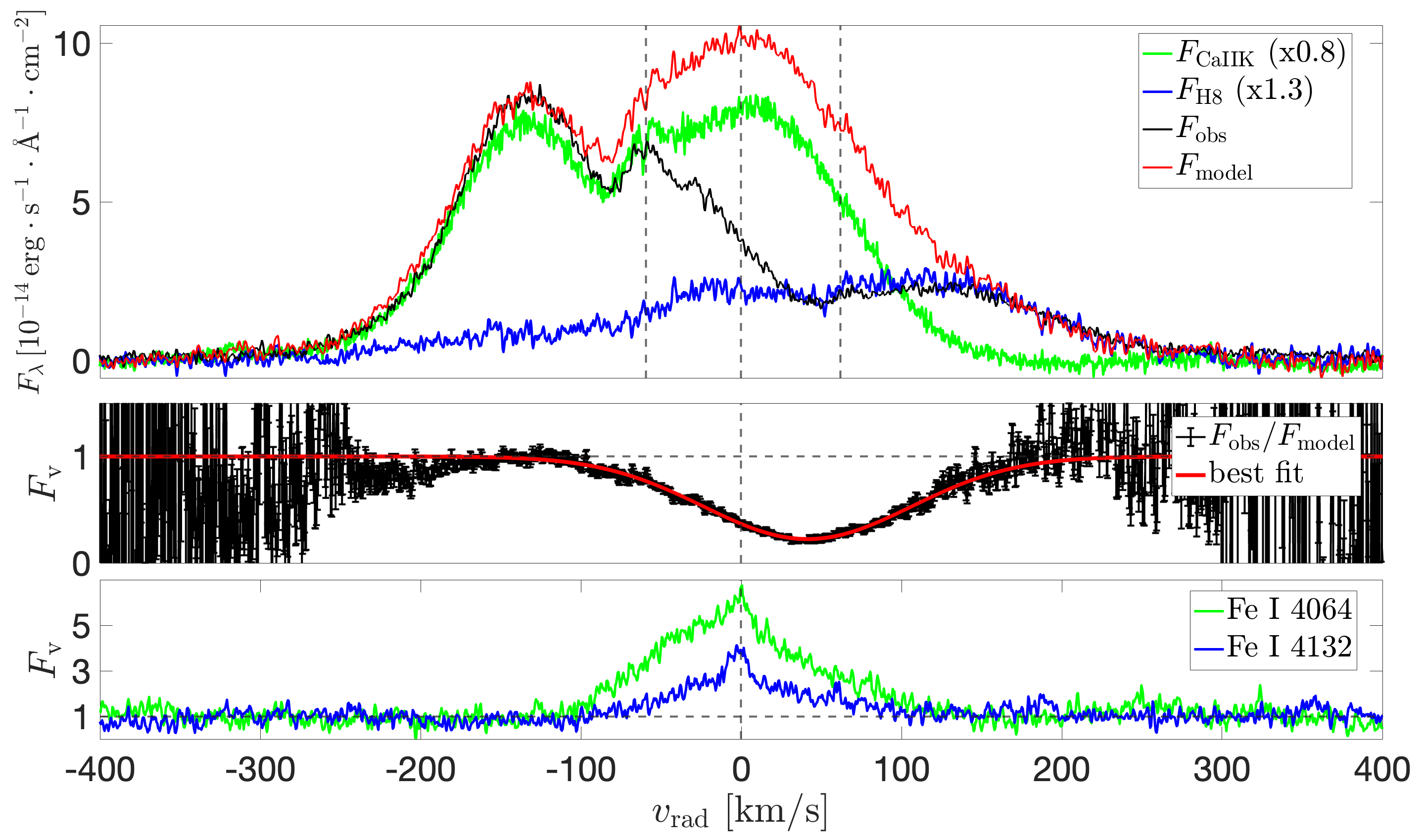}
		\caption{Reconstruction of the \ion{Fe}{i} absorption. All profiles are from the epoch 2 ESPRESSO spectrum. All lines are plotted relative to the rest wavelength of \ion{Fe}{i} 3969. 
		The top panel shows the continuum subtracted profile of the \ion{Ca}{ii}~H~+~H$\epsilon$ blend (black) compared to the \ion{Ca}{ii}~K (green) and H8 (blue) lines. The red line is the model for the unabsorbed blend (Eq.~\ref{blend_model}). The vertical dashed lines indicate the velocity displacement of \ion{Ca}{ii}~H, \ion{Fe}{i}~3969 and H$\epsilon$ relative to the rest wavelength of \ion{Fe}{i}~3969.
		The middle panel shows the \ion{Fe}{i} $3969$ absorption profile, reconstructed as explained in Sect.~\ref{FeI_fluorescence}, and its best fit.
		The bottom panel shows the \ion{Fe}{i} $4064$ and $4132$ emission lines.
		}
		\label{fig:FeI_3969}
	\end{figure}
    
    The third plot in the second row of Fig.~\ref{fig:ESPRESSO_LINES} compares the \ion{Fe}{i} $4064$, $4132$, and $4144$ lines. The first two are a doublet, having a common upper level.
    Although these three lines are from the same multiplet, the integrated flux in the \ion{Fe}{i}~$4132$ line, the weaker line of the doublet, is more than twice the integrated flux of the \ion{Fe}{i}~$4144$ line, as shown in Table~\ref{tab:metallic_lines_properties}.
    This behavior was first recognized by \citet{Herbig1945}, who proposed that the unusual strength of this doublet is due to a fluorescence mechanism. The \ion{Fe}{i} line at $3969.26$~{\AA}, which shares the upper level with the doublet, is nearly coincident in wavelength with \ion{Ca}{ii}~H ($3968.47$~{\AA}) and H$\epsilon$ ($3970.08$~{\AA}). Absorption in this line can produce an overpopulation of the $\rm{y^3F^0} ~ (\rm{J = 3})$ level, from which the fluorescent emission lines originate. 
    The top panel of Fig.~\ref{fig:FeI_3969} shows that the \ion{Ca}{ii}~H~$+$~H$\epsilon$ blend is strongly absorbed in the region of \ion{Fe}{i}~3969. 
    
    We reconstructed the \ion{Fe}{i} 3969 absorption profile by creating a model for the unabsorbed emission in the \ion{Ca}{ii}~H~+~H$\epsilon$ blend. To this end, we used the \ion{Ca}{ii} K and H8 lines as template for \ion{Ca}{ii} H and H$\epsilon$, respectively. We subtracted the continuum in these lines and shifted them in radial velocity to the position of \ion{Ca}{ii} H and H$\epsilon$ relative to the \ion{Fe}{i} 3969 rest wavelength. The nonabsorbed model profile was computed as a linear combination,
    \begin{equation}
        F_{\rm{model}} = 0.8 F_{\rm{CaIIK}} + 1.3 F_{\rm{H8}}.
        \label{blend_model}
    \end{equation}
    The coefficients were determined by matching the wings of the blended profile ($\lvert v_{\rm{rad}} \rvert \gtrsim 150~\rm{km~s^{-1}}$), where absorption from \ion{Fe}{i} is not expected, to the wings of the two template lines. The absorption profile was then obtained by dividing the observed \ion{Ca}{ii}~H + H$\epsilon$ blend by its model. The result is shown in the middle panel of Fig.~\ref{fig:FeI_3969}.
    We fit this profile with a Gaussian
    \begin{equation}
        F_{v} = 1 + C ~ \exp{\left[\frac{(v-v_0)^2}{2\sigma^2}\right]}.
    \end{equation}
    The best fit parameters are $C = -0.777 \pm 0.002$, $v_0 = 41.4 \pm 0.2~\rm{km~s^{-1}}$, $\sigma = 64.8 \pm 0.3~\rm{km~s^{-1}}$.
    The line width is in agreement with those observed in the fluorescent lines (bottom panel of Fig.~\ref{fig:FeI_3969}).
    
    The \ion{Fe}{i} 3969 profile provides constraints on the structure of the circumstellar envelope \citep{Willson1975}. Since the \ion{Ca}{ii} and \ion{H}{i} lines are formed in the pre-shock region (Sect.~\ref{HI_CaII_HeI}), a foreground structure that is opaque in \ion{Fe}{i} must be present to produce absorption. From the Gaussian best fit of the profile and assuming a purely extinguishing medium for which $F(v) = \exp(-\tau_v)$, we obtain a minimum optical depth at line center $\tau_0 \sim 1.5$. Therefore, this region is optically thick in \ion{Fe}{i}. This is confirmed by the integrated flux ratio of the fluorescent lines, $r_{\rm{FeI}}$, that is equal to $2.6$ at epoch 2, much lower than the ratio of the $A_{\rm{ji}}$ values, $5.6$. The line center is redshifted by $\sim 40~\rm{km~s^{-1}}$, indicating that the absorption is produced in infalling material. 
    
\begin{figure}
		\centering
		\includegraphics[width=\linewidth]{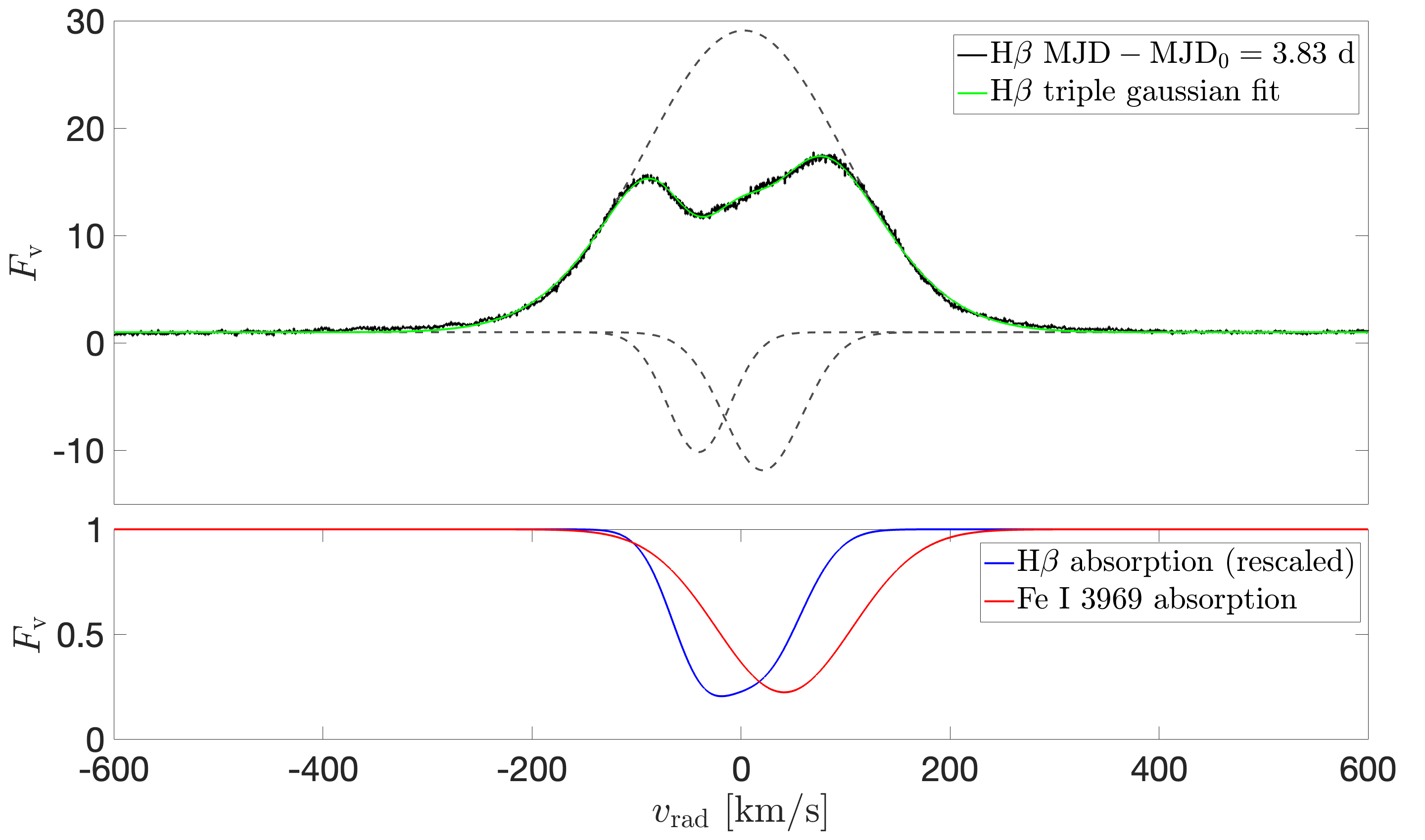}
		\caption{Comparison between the H$\beta$ and Fe I 3969 absorption components. Top panel: triple Gaussian fit of the continuum normalized H$\beta$ line in the epoch 2 ESPRESSO spectrum. The dashed lines show the three Gaussian components. Bottom panel: comparison between the continuum normalized absorption profiles in H$\beta$ and \ion{Fe}{i} $3969$. The red line is the Gaussian best fit of the \ion{Fe}{i} $3969$ from Fig.~\ref{fig:FeI_3969}. The blue line is the sum of the two Gaussian absorption components in H$\beta$, rescaled relative to the continuum to allow the comparison with the \ion{Fe}{i} $3969$ absorption profile.}
		\label{fig:Hbeta_vs_FeI_abs}
	\end{figure}

    Overall, the reconstructed \ion{Fe}{i}~3969 absorption profile is similar to the core of the Balmer lines (e.g., H$\beta$ in Fig.~\ref{fig:ESPRESSO_LINES}) in the epoch 1 and 2 ESPRESSO spectra. To compare the two features quantitatively, we fit the H$\beta$ line in the epoch 2 ESPRESSO spectrum with a triple Gaussian model. 
    We forced one Gaussian to fit the broad emission and the other two to reproduce the asymmetric depression. Figure~\ref{fig:Hbeta_vs_FeI_abs} displays the results of the fit. Compared to \ion{Fe}{i} $3969$, the absorption in H$\beta$ has a blueshifted centroid and extends to lower positive velocities. The difference probably stems from the fact that H$\beta$ is self-absorbed, while \ion{Fe}{i} $3969$ absorbs against \ion{Ca}{ii} H and H$\epsilon$. This means that, in the case of the Balmer lines, absorption can take place only if emission at a given radial velocity is intercepted by hydrogen gas moving at the same $v_{\rm{rad}}$\footnote{The difference can be on the order of the local thermal velocity.} along the line of sight. On the other hand, the fluorescence phenomenon couples regions having radial velocity differences equal to the shift between the rest wavelengths of \ion{Ca}{ii}~H and H$\epsilon$ relative to \ion{Fe}{i}~$3969$, that is, $\sim -59~\rm{km~s^{-1}}$ and $\sim +62~\rm{km~s^{-1}}$ respectively.
    This indicates the presence of velocity gradients in the accretion flow, in agreement with the magnetospheric accretion scenario. An outer zone of the magnetosphere sees \ion{Ca}{ii} H emission that is redshifted, leading to the tuned absorption in \ion{Fe}{i} $3969$ \citep{Gahm2001}. 
    For the same reason, absorption in \ion{Fe}{i} can take place due to \ion{H}{i} emission that is seen as blueshifted by \ion{Fe}{i}.  
    This result also demonstrates that the \ion{Ca}{ii} emission at moderate blueshifted velocities ($-100 \, \rm{km~s^{-1}} \lesssim v_{\rm{rad}} \lesssim 0 \, \rm{km~s^{-1}}$) is produced in the infalling gas.
    
	\subsection{Other metallic lines}
    \label{metallic_lines}
    \begin{table*}
    	\centering
    	\caption{Atomic parameters and integrated fluxes of the selected set of metallic lines. The spectra were dereddened with $A_{\rm{V}} = 1$~mag (Sect.~\ref{slab_model}) for the flux integration.} 
    	\begin{tabular}{rcccccccc}
    	    \hline
    	    Ion & $\chi_{\rm{I}} ~[\rm{eV}]$ & $\lambda ~ [${\AA}$]$ & A$_{\rm{ji}} ~ \rm{[s^{-1}]}$ & E$_{\rm{i}} ~[\rm{eV}]$ & E$_{\rm{j}} ~[\rm{eV}]$ & $g_{\rm{j}}$ & \multicolumn{2}{c}{F$_{\rm{int}} ~ [10^{-14} ~ \rm{erg ~ s^{-1} ~ cm^{-2}}]$} \\
    	    &  &  &  &  &  &  & Epoch $2$ & Epoch $3$ \\
    	    \hline
    	    \ion{Na}{I} & $5.14$ & $5889.95$ & $6.16 \cdot 10^7$ & $0.00$ & $2.10$ & $4$ & $5.8 \pm 0.5$ & $2.6 \pm 0.6$ \\
    	    \ion{Ca}{I} & $6.11$ & $4226.73$ & $2.18 \cdot 10^8$ & $0.00$ & $2.93$ & $3$ & $3.1 \pm 0.9$ & $1.3 \pm 1.2$ \\
    	    \ion{Mg}{I} & $7.65$ & $5183.60$ & $5.61 \cdot 10^7$ & $2.89$ & $5.11$ & $3$ & $4.5 \pm 0.7$ & $1.6 \pm 0.8$ \\
    	    \ion{Fe}{I} & $7.90$ & $4143.87$ & $1.33 \cdot 10^7$ & $1.56$ & $4.55$ & $9$ & $1.0 \pm 0.7$ & $0.5 \pm 1.0$ \\
    	    \ion{Fe}{I} & $7.90$ & $4063.59$ & $6.65 \cdot 10^7$ & $1.56$ & $4.61$ & $7$ & $7.2 \pm 1.6$ & $2.4 \pm 2.2$ \\
    	    \ion{Fe}{I} & $7.90$ & $4132.06$ & $1.18 \cdot 10^7$ & $1.61$ & $4.61$ & $7$ & $2.8 \pm 1.6$ & $0.3 \pm 0.6$ \\
    	    \ion{Fe}{I} & $7.90$ & $4404.75$ & $2.75 \cdot 10^7$ & $1.56$ & $4.37$ & $9$ & $1.8 \pm 1.0$ & $0.3 \pm 1.3$ \\
    	    \ion{Fe}{II} & $16.20$ & $4923.93$ & $4.30 \cdot 10^6$ & $2.89$ & $5.41$ & $4$ & $12.5 \pm 1.0$  & $5.3 \pm 1.2$ \\
    	    \ion{Fe}{II} & $16.20$ & $5316.61$ & $3.90 \cdot 10^5$ & $3.15$ & $5.48$ & $10$ & $9.6 \pm 0.7$ & $3.7 \pm 0.9$ \\
    	    \hline
    	    \ion{Ca}{II} & $11.87$ & $8542.09$ & $9.90 \cdot 10^6$ & $1.70$ & $3.15$ & $4$ & $7.67 \pm 1.1$ & $-$ \\
    	    \ion{Mg}{II} & $15.04$ & $4481.13$ & $2.33 \cdot 10^8$ & $8.86$ &  $11.63$ & $8$ & $< 0.1$ & $-$ \\
    		\hline
    	\end{tabular}
    	\label{tab:metallic_lines_properties} 
    \end{table*}
    Among the numerous metallic lines observed in the spectrum of HM~Lup, we selected a set of emission lines that belong to different species and have different ionization potentials and excitation energies, so that they potentially probe different conditions in the accretion structure. Table~\ref{tab:metallic_lines_properties} reports the atomic parameters of the selected transitions. The lines are shown in the second row of Fig.~\ref{fig:ESPRESSO_LINES}. All the lines display a NC+BC structure. The BC has a full width at half maximum (FWHM) of $ \sim 115~\rm{km~s^{-1}}$ and an emission profile that is skewed to the red. 
    The last two columns show the integrated flux for each line in the epoch 2~and~3 ESPRESSO spectra, chosen as representative of the line behavior during and just after the accretion burst. Before integrating, we dereddened the spectra using $A_{\rm{V}} = 1$~mag (Sect.~\ref{slab_model}) and the \citet{Cardelli+1989} extinction law with $R_{\rm{V}} = 3.1$.
    At epoch~2, the emission is strongest in \ion{Fe}{ii}, followed by the \ion{Na}{i}~D lines. 
    \ion{Fe}{i} lines are weaker than \ion{Fe}{ii} lines, typically by a factor of $\sim 5-7$, except for the \ion{Fe}{i} $4064$ and $4132$ lines (Sect.~\ref{FeI_fluorescence}). This is different than EX~Lup, for which \citet{Sicilia-Aguilar+2012} found a ratio of two for the \ion{Fe}{ii} vs. \ion{Fe}{i} during the outburst. 
	
	\begin{figure}
		\centering
		\includegraphics[width=\linewidth]{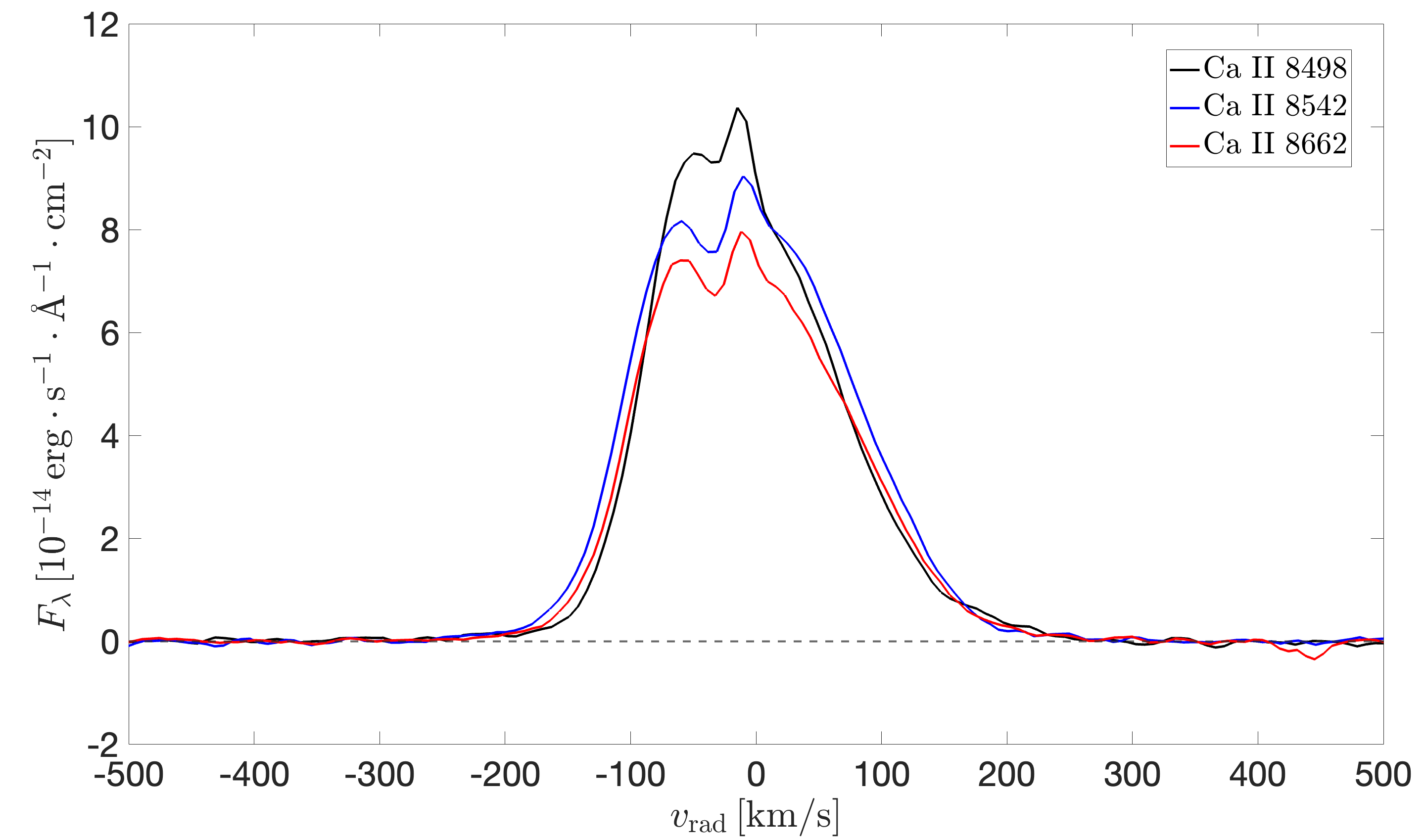}
		\caption{Continuum subtracted profiles of the \ion{Ca}{ii} IRT lines in the X-Shooter spectrum at MJD - MJD$_0 = 3.80$~d, almost simultaneous to the ESPRESSO epoch 2 spectrum.}
		\label{fig:CaII_IRT}
	\end{figure}
    
    Figure~\ref{fig:CaII_IRT} shows the \ion{Ca}{ii} infrared triplet (IRT) lines from the 2021 X-Shooter spectrum, that have profiles similar to those of the metallic lines. The components of the \ion{Ca}{ii} IRT have relative intensities that differ from the 1:9:5 relation that is expected from the ratios of their gf-values\footnote{The gf-values are proportional to $g_{\rm{j}} \cdot A_{\rm{ji}}$.}, indicating that these lines are formed in an optically thick environment. This sets a lower limit to the electron density of $10^{11} ~ \rm{cm}^{-3}$ \citep{HamannPersson1992}.
    
    Additional information on the optical depth of the medium can be obtained from the flux ratio of lines that share the upper level. We chose the \ion{Fe}{ii}~4549 and 4352 lines ($E_{\rm{j}} = 5.55$~eV) for this exercise, since they are strong, not blended and near in wavelength. 
    The integrated flux ratio for the epoch 2 ESPRESSO spectrum is $r_{\rm{FeII}} \approx 1.16$. This value is lower than the ratio of their $A_{\rm{ji}}$ values, $2.04$, confirming that the region where the metallic lines are formed is optically thick. 
    This complicates the derivation of the properties of the medium in terms of line ratios, since the line emission depends on the escape probability \citep{Sobolev1960, KogureLeung2007}. 
    The comparison between the line ratio and the ratio of the $A_{\rm{ji}}$ values for the selected \ion{Fe}{i} and \ion{Fe}{ii} doublets indicate that this region is more optically thick in \ion{Fe}{i} than in \ion{Fe}{ii}. Moreover, the observation of the \ion{Fe}{i} fluorescence (Sect.~\ref{FeI_fluorescence}) suggests that non-LTE (NLTE) effects can be important in the region where the metallic lines are formed. 
    Therefore, the \ion{Fe}{ii} to \ion{Fe}{i} ratio cannot be taken as an unambiguous diagnostic of the actual conditions of this region. 
    
    Some constraints on $n_{\rm{e}}$ and $T$ can still be placed from line ratios under the assumption of LTE conditions (Appendix~\ref{LTE_eqs}). In the conditions for the coexistence of \ion{Mg}{i}, \ion{Fe}{i} and \ion{Fe}{ii}, neutral sodium and calcium are almost completely ionized. In such an environment, the \ion{Na}{i}~D and \ion{Ca}{i} $4227$ lines are therefore less optically thick than the \ion{Fe}{i} lines.
    Hence, the \ion{Ca}{ii} to \ion{Ca}{i} ratio is a better temperature indicator than the \ion{Fe}{ii} to \ion{Fe}{i} ratio. The \ion{Ca}{i} $4227$ line can be compared to one of the IRT lines, for instance the \ion{Ca}{ii} $8542$ line.
    Another temperature diagnostic for the emitting region is the \ion{Mg}{ii} to \ion{Mg}{i} ratio. The non-detection of \ion{Mg}{ii} $4481$, which has an upper state with $E_{\rm{j}} = 11.63$~eV, provides an upper limit to the temperature. 
    We computed this upper limit by estimating the noise level in the continuum subtracted spectrum at the nominal position of the \ion{Mg}{ii} line. The atomic parameters of the \ion{Mg}{ii} $4481$ line and the \ion{Ca}{ii} $8542$ line are reported in the last two rows of Table~\ref{tab:metallic_lines_properties}. Figure~\ref{fig:Saha_metals} shows the result of the LTE analysis described in Appendix~\ref{LTE_eqs} for Ca and Mg. For $11 \lesssim \log n_{\rm{e}} \lesssim 14$, the \ion{Ca}{ii} to \ion{Ca}{i} ratio is reproduced with temperatures between $\sim 4000$ and $7000$~K, in agreement with the density-dependent upper limit on $T$ placed by the nondetection of \ion{Mg}{ii}~$4481$ (Fig.~\ref{fig:Saha_metals}). 
    
     \begin{figure}
		\centering
		\includegraphics[width=\linewidth]{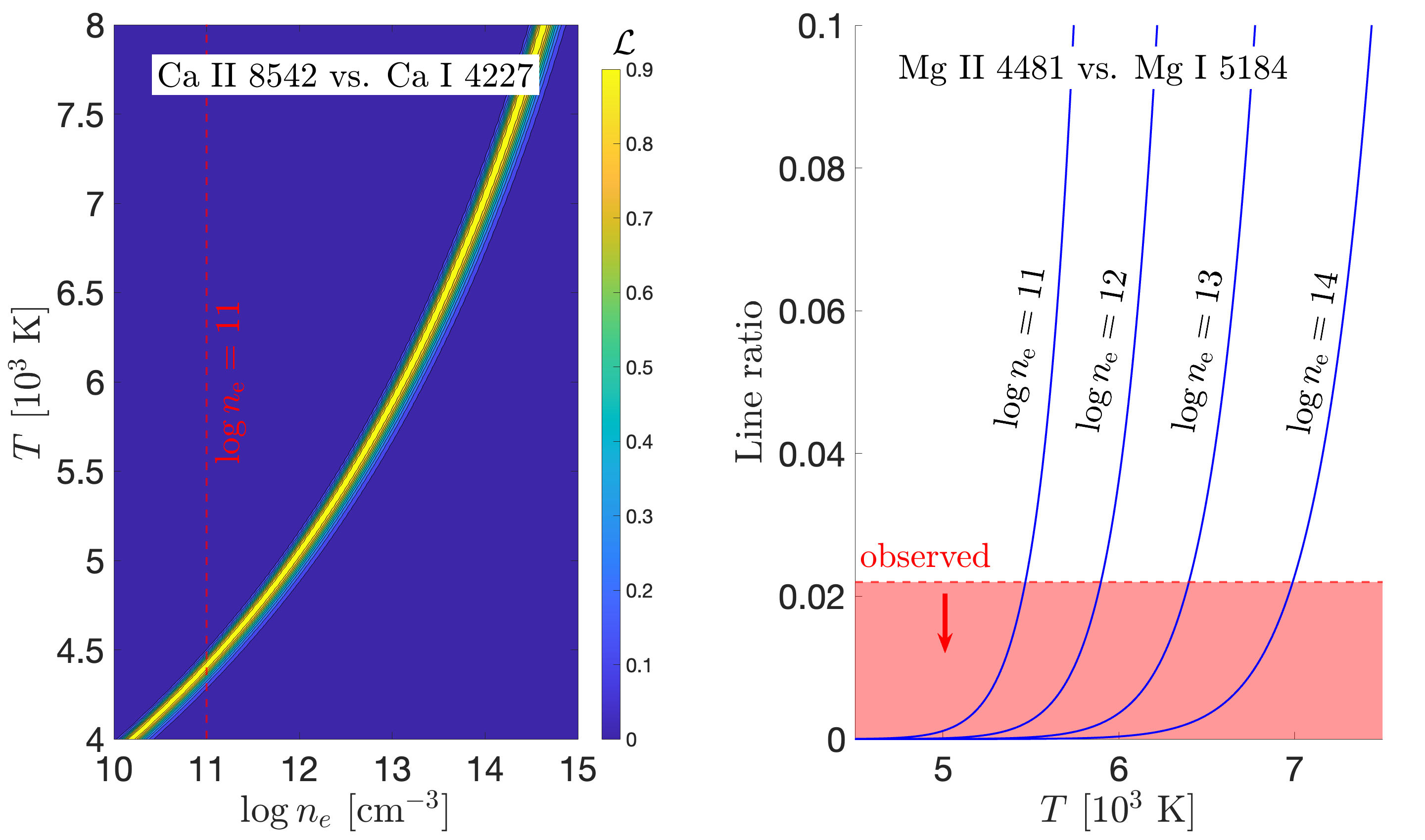}
		\caption{Result of the Saha-Boltzmann analysis for Ca and Mg. Observed line ratios were computed from the epoch~2 ESPRESSO spectrum. Left panel: likelihood for the observed \ion{Ca}{ii} $8542$ vs. \ion{Ca}{i} $4227$ ratio as a function of $\log n_{\rm{e}}$ and $T$. Right panel: theoretical \ion{Mg}{ii} $4481$ vs. \ion{Mg}{i} $5184$ ratios as a function of $T$ for $\log n_{\rm{e}} ~ [\rm{cm}^{-3}] = 11, 12, 13, 14$. The red area marks the upper limit on the observed ratio.}
		\label{fig:Saha_metals}
	\end{figure}
	
    \section{Spectrophotometric variability}
    \label{spectrophotometry}
    
    \subsection{Photometry}
	\label{sect:photometry}
	Figure~\ref{fig:simultaneous_spectrophot} summarizes all available photometric data in 2021, while Fig.~\ref{fig:TESS_2019} shows the TESS Sector\,12 light curve from May 2019.
	We lack AAVSO data in the highest flux peak of the 2021 TESS light curve, that is, in the range $3~\rm{d}~\lesssim~\rm{MJD - MJD_0}~\lesssim~5$~d, which is coincident with the X-Shooter spectrum. To complement the multiband light curve at that epoch, we computed synthetic $BVR_{\rm{c}}I_{\rm{c}}$  photometry from the X-Shooter spectrum using the filter transmission curves downloaded from the SVO Filter Profile Service\footnote{\url{http://svo2.cab.inta-csic.es/theory/fps/}} \citep{SVO_2012, SVO_2020}.
	
	There are differences in the light curve morphology between the two TESS epochs. The 2019 TESS data show no clear periodicity pattern and a tendency for showing dips, while quasi-periodic spikes are observed in the 2021 TESS photometry. 
	To quantify the different morphology, we computed the variability metrics introduced by \citet{Cody+2014} for both light curves. The parameter $Q$ describes the degree of periodicity and the parameter $M$ the (a)symmetry around the median value of a given light curve. $Q$ ranges between $0$ (= highly periodic) and $1$ (= aperiodic), while strongly negative (positive) values of $M$ indicate the tendency of showing bursts (dips) in the light curve. 
    We obtained $Q = 0.39$ and $M = -0.66$ for the 2021 TESS light curve and $Q = 0.57$ and $M = -0.18$ for the 2019 TESS light curve, suggesting a higher degree of periodicity and a more pronounced bursting behavior in 2021.  
    
	We performed a Lomb-Scargle \citep{Lomb1976, Scargle1982} analysis on the TESS light curves. The Lomb-Scargle Periodograms (LSPs) for the TESS 2021 and 2019 light curves are shown in the insets of Figs.~\ref{fig:simultaneous_spectrophot} and~\ref{fig:TESS_2019}, respectively.
	We detected four peaks in both LSPs, all below a False Alarm Probability (FAP) of $10^{-5}$. We chose the periods associated to the highest peak of each periodogram, $4.79$~d and $5.41$~d for the light curves from 2021 and 2019 respectively, as representative of the periodicity of the light curves. This is shown in Figs.~\ref{fig:simultaneous_spectrophot} and~\ref{fig:TESS_2019} by visually comparing the light curves to an average profile produced by phase folding the data with those periods and smoothing the phase folded light curve with a boxcar filter with a width of $0.25$ in phase, as described by \citet{Cody+2014}.
	We estimated the uncertainties on the periods as the standard deviation of a Gaussian function fitted to the peaks in the LSPs. The results are $P = 4.79 \pm 0.28$~d for the year 2021 and $P = 5.41 \pm 0.35$~d for 2019. Although the central value is different, the two periods agree within the uncertainties. The LSP of both TESS light curves present additional, weaker peaks with the same structure. For the case of the 2021 light curve they are at $3.76$, $6.20$ and $9.58$~d. The 2019 light curve shows peaks at similar values. The latter is the first harmonic of the $4.79$~d period. 
	To understand whether the other two periods are significant, we produced an average profile analogous to the one shown in purple in Fig.~\ref{fig:simultaneous_spectrophot}. The comparison to the observed light curve showed poor match. This suggests that these peaks are aliases.
	
	
	\begin{figure*}
		\centering
		\includegraphics[width=\linewidth]{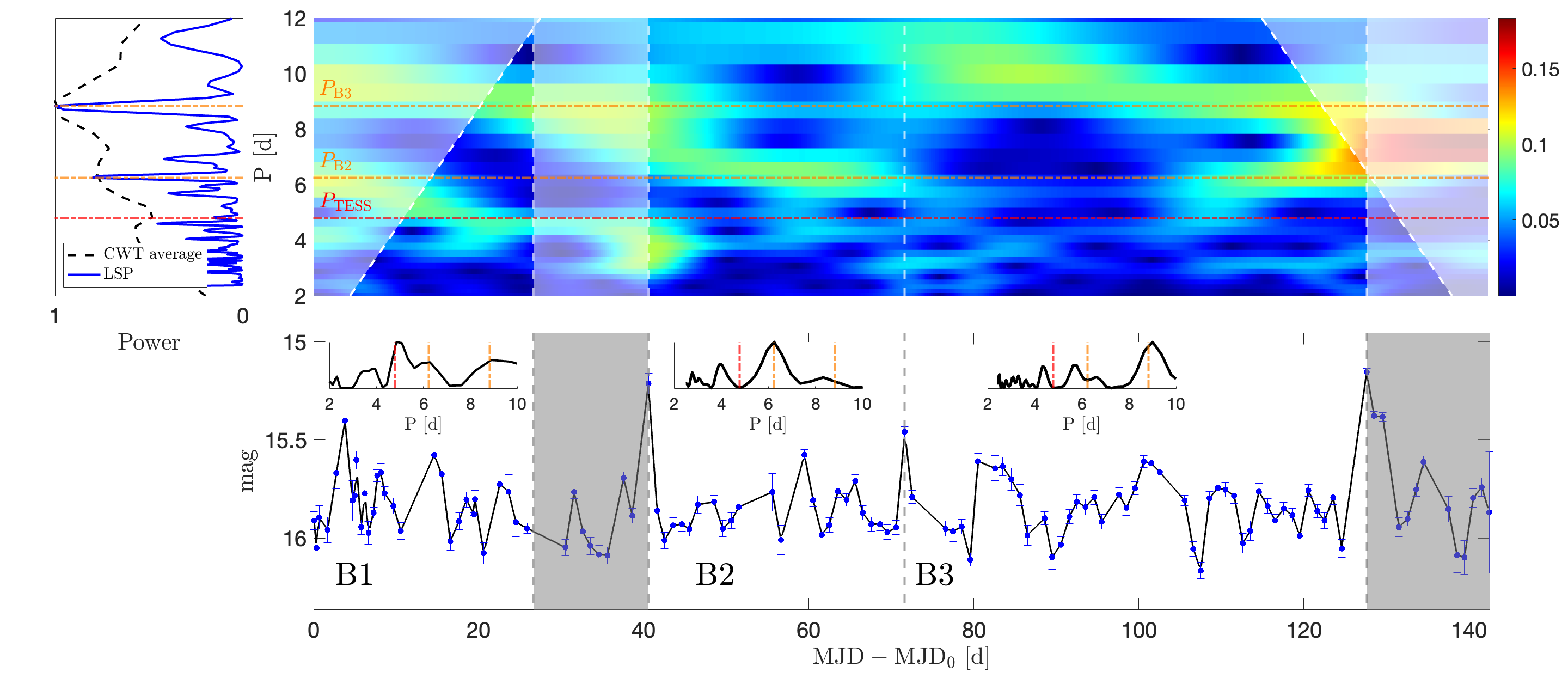}
		\caption{Continuous wavelet analysis of the AAVSO $B$ band photometry. The top panel shows the CWT. The inset on the left displays the LSP of the whole dataset, together with a temporal average of the CWT. The orange horizontal dot-dashed lines mark the periods obtained from the LSP analysis, i.e., $P_{\rm{B2}} = 6.24$~d and $P_{\rm{B3}} = 8.84$~d, while the red one indicates the 2021 TESS period, $P_{\rm{TESS}} = 4.79$~d. The bottom panel displays the AAVSO $B$ band light curve and its linear interpolation. The insets show the LSP for three different time segments. We excluded the shaded areas from the calculation of the LSPs to highlight the different timescales detected. 
		}
		\label{fig:Bmag_CWT} 
	\end{figure*}

	We studied the periodicity of the AAVSO light curves by means of continuous wavelet analysis, which is a useful tool to determine the frequency content of a signal as a function of time.
	We chose as wavelet template the Morse wavelet \citep{LillyOlhede2012} with a symmetry parameter $\gamma = 3$ and a time-bandwidth product $\mathcal{P} = 90$, that produces a time resolution of $\sim 30$~d. Since the wavelet analysis works only on evenly spaced data, we linearly interpolated the AAVSO data to $0.25$~d centers to remove the gaps in the light curves. Higher cadences did not change the results of the analysis. 
	The continuous wavelet transform (CWT) of the $B$ band photometry is shown in Fig.~\ref{fig:Bmag_CWT}. The light curve and its CWT can be divided into three time segments with different power spectra, as shown in the lower panel of Fig.~\ref{fig:Bmag_CWT}. In the first segment (B1), which is simultaneous to the 2021 TESS light curve, the detected period $P_{\rm{B1}}$ is compatible with the period of the TESS light curve, $4.79$~d. In the other two segments the main periods detected are $P_{\rm{B2}} = 6.24$~d and $P_{\rm{B3}} = 8.84$~d.
	In the LSP of the whole light curve, in which the time dependence of the power spectrum is lost, these multiple contributions are averaged out. We considered only $P_{\rm{B3}}$ as statistically relevant, since it has $\rm{FAP} = 2.5\%$ in the LSP.
	
	The part of the AAVSO data that is simultaneous with TESS data follow the shape of the TESS light curve, as shown in Fig.~\ref{fig:simultaneous_spectrophot}. The relative magnitude variations vary across the bands, being $0.3$~mag in $R_{\rm{c}}$ and $I_{\rm{c}}$, $0.5$~mag in $V$ and $0.7$~mag in $B$. The approximately enhanced variability amplitude in the blue part of the spectrum is compatible with changes in the accretion rate, since the excess flux is expected to peak at shorter wavelengths \citep{CalvetGullbring1998}. 

	\subsection{Accretion rate variability}
	\label{slab_model}
	
	\begin{figure}
		\centering
		\includegraphics[width=\linewidth]{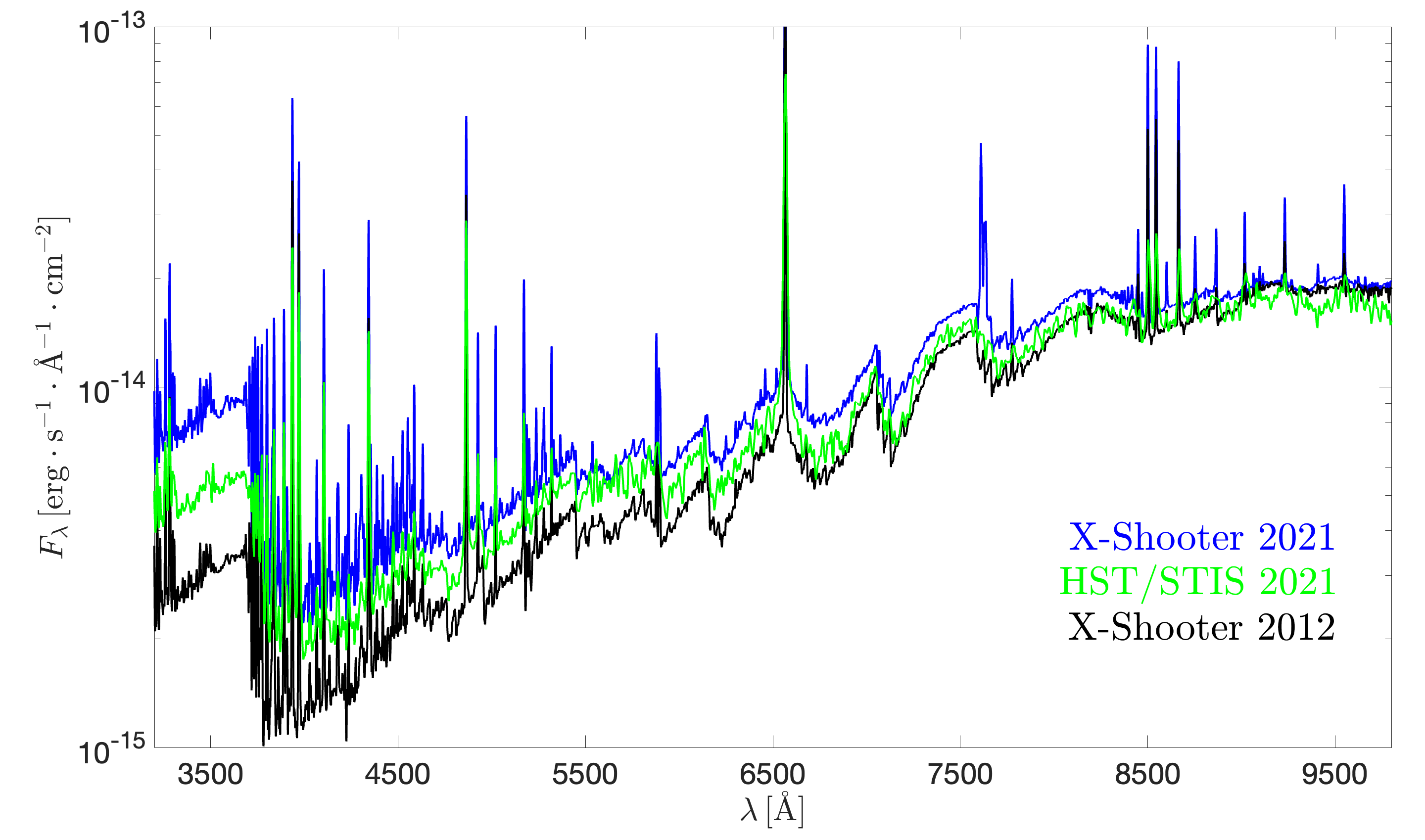}
		\caption{VLT/X-Shooter and HST/STIS spectra of HM~Lup in the STIS wavelength range. The spectra were smoothed to the HST resolution with a Gaussian filter for clarity.
		}
		\label{fig:XSHOOT_vs_HST}
	\end{figure}
	
	\begin{table*}
    	\centering
    	\caption{Main accretion parameters derived from the spectra.
    	} 
    	\begin{tabular}{lccccccc}
    	    \hline
    	    Instrument & $\rm{MJD-MJD_0}$~[d] & $A_{\rm v}$~[mag]& \multicolumn{2}{c}{$\log{(L_{\rm{acc}}/L_{\odot})}$} & \multicolumn{2}{c}{$\dot{M}_{\rm{acc}} ~ [10^{-9}~\rm{M_{\odot}~yr^{-1}}]$} & $r_{6000}$ \\
    	     & & & Slab Model & Lines & Slab Model & Lines  \\
    	    \hline
            X-Shooter   & $-3298.17$ & $0.75$ & $-1.77 \pm 0.25$ & $-1.50 \pm 0.07$ & $2.5$ & $4.8$ & $\leq 0.2\ddagger$  \\
    		ESPRESSO    & $2.89$ & $1^{\dagger}$ & $-$ & $-1.16 \pm 0.08$ & $-$ & $10.5$ & $2.24 \pm 0.76$   \\
    		X-Shooter   & $3.80$ & $1$ & $-1.20 \pm 0.25$ & $-1.08 \pm 0.07$ & $9.5$ & $12.7$ & $1.60 \pm 0.2\ddagger$  \\
    		ESPRESSO    & $3.83$ & $1^{\dagger}$ & $-$ & $-1.09 \pm 0.07$ & $-$ & $12.4$ & $1.41 \pm 0.74$ \\
    		HST/STIS    & $5.14$ & $1^{\dagger}$ & $-1.37 \pm 0.25$ & $-1.23 \pm 0.08$ & $6.4$ & $9.0$ & $-$ \\
    		ESPRESSO    & $5.75$ & $1^{\dagger}$ & $-$ & $-1.27 \pm 0.08$ & $-$ & $8.2$ & $0.92 \pm 0.44$ \\
    		ESPRESSO    & $12.86$ & $1^{\dagger}$ & $-$ & $-1.21 \pm 0.08$ & $-$ & $9.4$ & $0.6 \pm 0.3$ \\
    		\hline
    	\end{tabular}
    	\label{tab:spectra_obs} 
    	\tablefoot{
    	$^{\dagger}$: extinction assumed from X-Shooter 2021 fit; $^{\ddagger}$: veiling at $6200 \textup{~\AA}.$
    	}
    \end{table*}
	
	A comparison between the two X-Shooter spectra and the STIS spectrum in the $3200-10000$~{\AA} range is shown in Fig.~\ref{fig:XSHOOT_vs_HST}. 
	Marked differences in the Balmer continuum and the Balmer jump are evident between the three spectra. In the 2021 X-Shooter spectrum the Balmer continuum was $\sim 3$ times higher than in 2012. The ratio is reduced to a factor $\sim 1.5$ in the Paschen continuum.
	The flux in the STIS spectrum lies approximately between these two. 
	
	We derived the accretion parameters by fitting each spectrum with a combination of a photospheric (nonaccreting) template and a model for continuum emission from a slab of hydrogen representing the accretion shock. This method yields simultaneously with the UV accretion luminosity ($L_{\rm acc,UV}$), the spectral type, and extinction ($A_{\rm V}$) of the star. For more details on the procedure we refer to \citet{Manara+2013}. 
	For the 2012 spectrum we used published values obtained with the same method. That spectrum was first analyzed by \citet{Alcala+2014}, but in this work we use the updated values from \citet{Manara+2022}. 
	The resulting values of $A_{\rm V}$, $L_{\rm acc,UV}$, and $\dot{M}_{\rm acc}$ are listed in Table~\ref{tab:spectra_obs}. Typical uncertainties on these parameters are $0.1$~mag, $0.25$, and $0.45$~dex respectively \citep{Manara+2013, Alcala+2014, Manara+2021}.
	The best fit of the 2021 X-Shooter spectrum returned a SpT M2, in agreement with the fit obtained by \citet{Alcala+2014} for the 2012 spectrum, and $A_{\rm{V}} = 1$~mag. 
	The STIS spectrum was fitted in the X-Shooter wavelength range. Its low resolution ($R \sim 1500$) prevents constraining the SpT. Therefore, we fixed it to M2 in the fitting routine and obtained $A_{\rm V} = 1.5$~mag and $\dot{M}_{\rm acc} = 1.53 \cdot 10^{-8}~\rm{M_{\odot}~yr^{-1}}$. Since this spectrum is only $\sim 1.3$~d separated in time from the second X-Shooter observation, variations in $A_{\rm V}$ on the order of $\Delta A_{\rm V} = 0.5$~mag seem unlikely, given that variations in the TESS bandpass are only about $0.2$\,mag, as shown in Fig.~\ref{fig:simultaneous_spectrophot}. 
	Hence, we fixed $A_V = 1$~mag in the STIS best fit, obtaining the accretion rate reported in Table~\ref{tab:spectra_obs}. 
	
	\begin{figure}
		\centering
		\includegraphics[width=\linewidth]{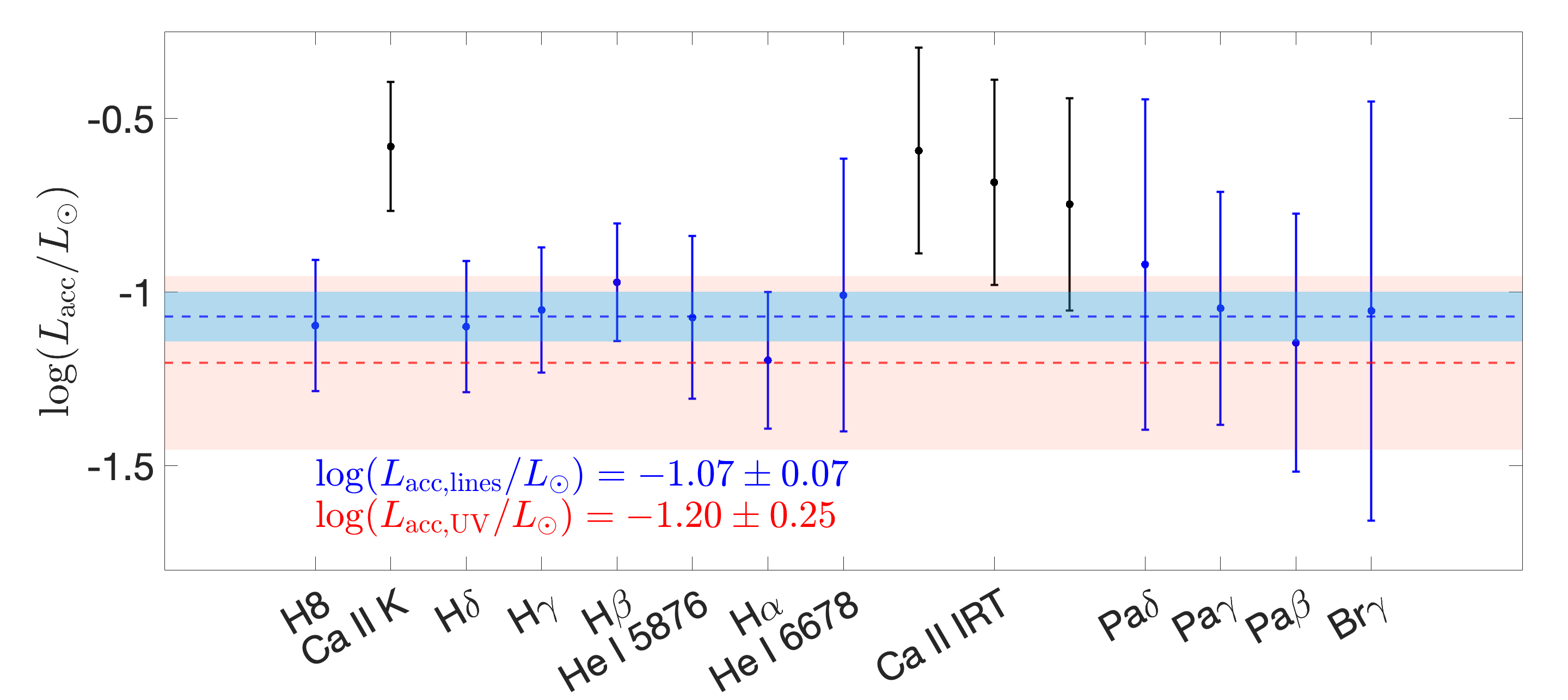}
		\caption{Comparison between the accretion luminosity derived from emission lines and that derived from the UV excess in the 2021 X-Shooter spectrum. $L_{\rm acc,UV}$ and its uncertainty are displayed with the red dashed line and the red shaded area. The blue dashed line and the blue shaded area are the value of $L_{\rm{acc,lines}}$ and its uncertainty, obtained from a weighted average of the line values. The lines marked in black were excluded from the calculation of the average.}
		\label{fig:XSHOOTER_Lacc}
	\end{figure}
	
	We obtained an independent measure of the accretion rate from the line luminosities, $L_{\rm{line}}$, using their empirical relation with $L_{\rm acc,UV}$, calibrated by \citet{Alcala+2017}. For the X-Shooter spectra, we integrated the available Balmer and Paschen lines, Br$\gamma$, \ion{Ca}{ii} K and infrared triplet (IRT) and \ion{He}{i} $5876$ and $6678$ lines. Since the ESPRESSO spectra cover a narrower wavelength range, we integrated only the Balmer, \ion{He}{i}, and \ion{Ca}{ii} K lines. For the STIS spectrum, we excluded the \ion{He}{i} lines because of their low S/N. Before integrating, we dereddened the spectra using the $A_{\rm V}$ value obtained from the slab model fit of the X-Shooter spectrum. Fig.~\ref{fig:XSHOOTER_Lacc} shows the comparison between the accretion luminosity derived from the lines, $L_{\rm acc,lines}$, and $L_{\rm{acc,UV}}$ for the 2021 X-Shooter spectrum. Most of the lines are in agreement with the accretion luminosity obtained from the slab model, but the \ion{Ca}{ii} lines overpredict the accretion luminosity. 
	For this reason, we excluded these lines from the calculation of $L_{\rm acc,lines}$. We further discuss this issue in Sect.~\ref{Tstrat}.
	A global value for $L_{\rm acc,lines}$ was then derived from a weighted average of the values obtained for the individual lines. This mean value was used to compute the accretion rate from the formula $\dot{M}_{\rm acc} = 1.25 L_{\rm acc} R_{\star}/(GM_{\star})$ \citep{Hartmann+1998}. The results are reported in Table~\ref{tab:spectra_obs}.
	
	Additional information on the variations in the accretion rate of CTTSs can be derived by measuring the veiling, that is, the excess emission due to the accretion process that makes photospheric absorption lines appear less deep \citep{Hartigan+1989}. This parameter is defined as the ratio of the excess flux relative to the photospheric flux, $r_{\lambda} = F_{\rm acc}(\lambda)/F_{\rm phot}(\lambda)$. We computed $r_{\lambda}$ at $6000$~{\AA} for the available spectra using the ROTFIT routine, as described by \citet{Frasca+2015} and \citet{Manara+2021}. The result is shown in the last column of Table~\ref{tab:spectra_obs}.
	
    The comparison between $L_{\rm acc}$ in 2012 and 2021 indicates that HM\,Lup was observed in a more active state in 2021, with an accretion rate $\sim 3-4$ times higher. The computed values of $\dot{M}_{\rm{acc}}$ and $r_{6000}$ support the hypothesis that the photometric behavior of the system in the main burst of the 2021 TESS light curve was caused by an increase in the accretion rate by a factor of $\sim 1.5$. 
    
    \subsection{Line variability}
    \label{line_variability}
    The variability of the H$\alpha$, H$\beta$, \ion{Ca}{ii} K, and \ion{He}{i} 5876 and 6678 lines between the four ESPRESSO spectra is shown in the first row of Fig.~\ref{fig:ESPRESSO_LINES}. The shape of the \ion{He}{i} lines is invariant, but the line flux changes. It is highest in the second epoch and lowest in the third and fourth, in agreement with the photometric behavior during the accretion burst. The H$\beta$ and \ion{Ca}{ii} K lines evolve from a double peaked structure in the first two epochs to a profile with a single red-shifted peak in the third and fourth. H$\alpha$ always remains flat-topped, suggesting that the double peaks are caused by an optical depth effect. If the temperature and density of the magnetosphere increase, as is likely to happen during the accretion burst, H$\beta$ and the higher lines of the series can develop a self-absorption component \citep{Muzerolle+2001}. The same happens for \ion{Ca}{ii} K which, being a resonance line, is more prone to absorption. 
    On the other hand the H$\alpha$ line, given its higher opacity, can be strongly optically thick and thermalized in these physical conditions, resulting in the disappearance of the self-absorption component \citep{Hartmann+2016}. 
    
    The flux emitted in the metallic lines follows the photometric behavior of the system. It is strongest in the first two epochs, while it returns to its quiescent level after the accretion burst, as suggested by the comparison of line fluxes between the third ESPRESSO epoch and 2012 X-Shooter spectra in Fig.~\ref{fig:ESPRESSO_LINES}. In the third and fourth epoch, the BC is absent in \ion{Fe}{i} and reduced in flux by a factor of $\sim 2-3$ in \ion{Fe}{ii}, \ion{Mg}{i}, and \ion{Na}{i}. The fact that the emission is stronger in both the \ion{Fe}{i} and \ion{Fe}{ii} lines during the high accretion state suggests that the outburst spectrum is mainly the result of a density enhancement, that is, an overall increase in the number of emitters, rather than a temperature increase. Table~\ref{tab:metallic_lines_properties} shows how the \ion{Fe}{ii} to \ion{Fe}{i} ratio at epoch 2 is lower than at epoch 3. This is compatible with an increase in the electron density during the outburst that favors the LTE population of \ion{Fe}{i} relative to \ion{Fe}{ii} (Appendix~\ref{LTE_eqs}).
    
\begin{figure}
		\centering
		\includegraphics[width=\linewidth]{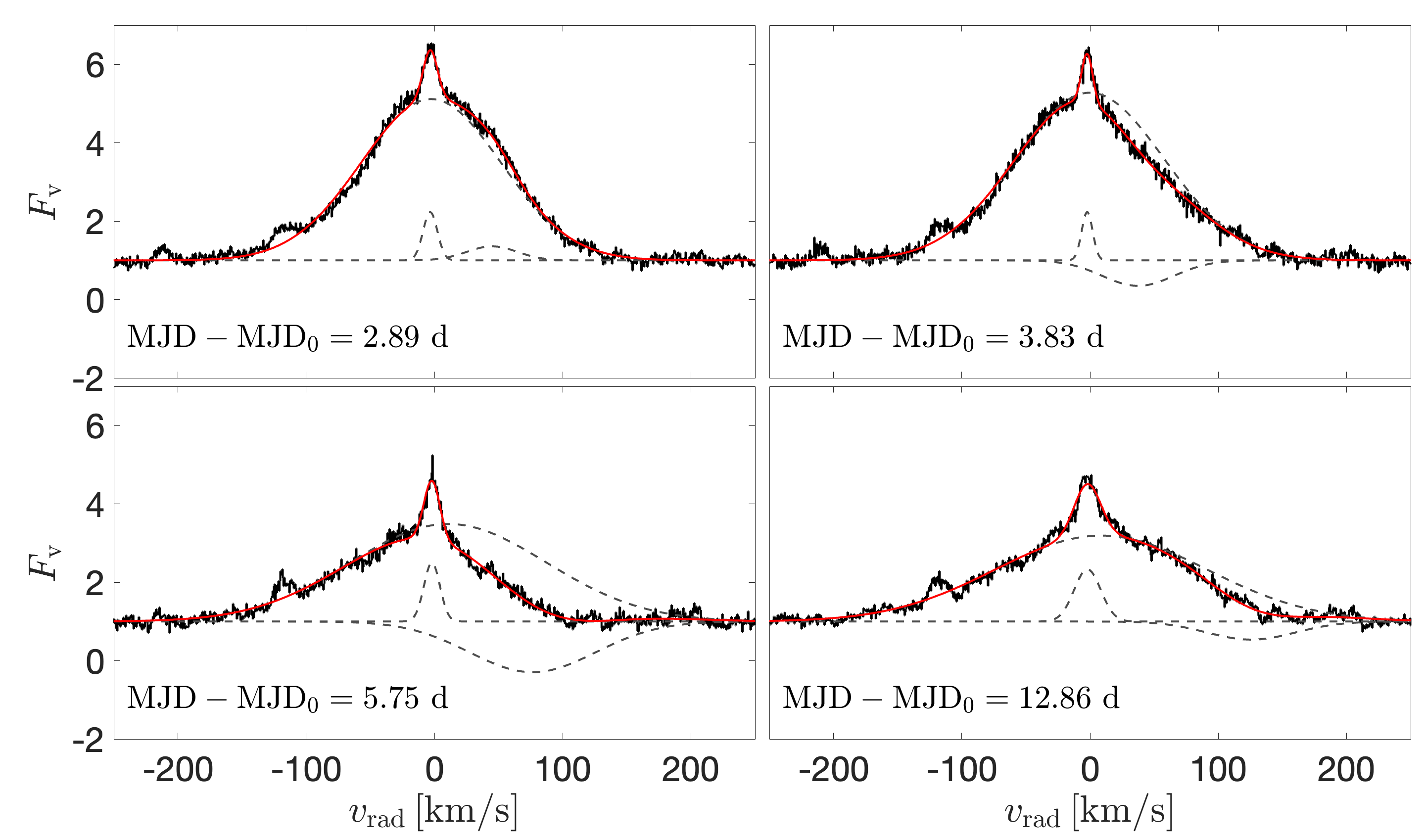}
		\caption{Triple Gaussian fit of the \ion{Fe}{ii} 4924 line profiles in the four ESPRESSO spectra. See Appendix~\ref{FeII4924_fit} for the fit function.}
		\label{fig:FeII_4924}
	\end{figure}

    To quantify the line variability we fit the \ion{Fe}{ii} 4924 line in each ESPRESSO epoch with a triple Gaussian model (Appendix~\ref{FeII4924_fit}) as shown in Fig.~\ref{fig:FeII_4924}. We chose this transition as template for the line variability because it has the highest S/N among the metallic lines. The best fit parameters are reported in Table~\ref{tab:FeII_4924_bestfit}.
    The line variations can be explained in terms of a broad emission component plus a variable redshifted absorption that reproduces the observed red skewness. The line is symmetric at epoch 1. Then, a redshifted absorption at $\sim +40~\rm{km~s^{-1}}$ appears at epoch 2, it becomes more pronounced, broader and at higher positive velocities at epoch 3, and eventually disappears at epoch 4. The evolution of the redshifted absorption is indicative of optical depth changes in the infalling material, which is likely associated to the rotation of a non-axisymmetric accretion structure, similar to what observed in EX\,Lup \citep{Sicilia-Aguilar+2012} or CVSO109 \citep{Campbell-White+2021}.
    

    \begin{table*}
    	\centering
    	\caption{Best fit parameters for the \ion{Fe}{ii} 4924 line profiles in the four ESPRESSO spectra. Since we are interested in fitting the BC only, we do not report the parameters for the NC, i.e., $C_1$, $v_1$, $\sigma_1$. 
    	} 
    	\begin{tabular}{cccccccc}
    	    \hline
    	    $\rm{MJD-MJD_0}$~[d] & $C_2$ & $v_2~\rm{[km~s^{-1}]}$ & $\sigma_2~\rm{[km~s^{-1}]}$ & $C_3$ & $v_3~\rm{[km~s^{-1}]}$ & $\sigma_3~\rm{[km~s^{-1}]}$ \\
    	    \hline
    	    $2.89$ & $4.12 \pm 0.05$ & $-2.6 \pm 0.7$ & $54.5 \pm 0.4$ & $0.36 \pm 0.07$ & $45 \pm 3$ & $20 \pm 4$ \\
    	    $3.83$ & $4.28 \pm 0.04$ & $-0.2 \pm 0.5$ & $57.9 \pm 0.2$ & $-0.65 \pm 0.05$ & $38 \pm 1$ & $28 \pm 2$ \\
            $5.75$ & $2.5 \pm 0.2$ & $11 \pm 5$ & $76 \pm 2$ & $-1.3 \pm 0.2$ & $77 \pm 2$ & $49 \pm 3$ \\
            $12.86$ & $2.19 \pm 0.02$ & $9 \pm 1$ & $82 \pm 1$ & $-0.46 \pm 0.04$ & $126 \pm 2$ & $35 \pm 2$ \\
    		\hline
    	\end{tabular}
    	\label{tab:FeII_4924_bestfit} 
    \end{table*}

    \subsection{Epoch 4 ESPRESSO spectrum}
    \label{ESPRESSO_4}
    Although the veiling is constant within the uncertainties between the epoch 3 and 4 ESPRESSO spectra, the accretion rate predicted from the lines is higher in epoch 4 than in epoch 3. This happens because the redshifted emission peak increases relative to the continuum in the higher lines of the Balmer series, as shown in Fig.~\ref{fig:ESPRESSO_LINES}. On the other hand, the \ion{He}{i} BC does not change between the two epochs. This further supports the hypothesis that the lines are formed in a stratified environment, both in temperature and in density. The Balmer emission peaks are not linked to the region that produces the veiling, namely, the continuum emission, unlike the wings of the Balmer and \ion{He}{i} lines. 
    
    Although we lack photometric information for the ESPRESSO epoch 4 spectrum, the phase-folded TESS light curve suggests that during this epoch the system is roughly in the same configuration as in the first two epochs. The phases are indeed $\phi_4 = 1.97$, $\phi_1 = -0.11$ and $\phi_2 = 0.08$.
    However, the line profiles are remarkably different. The metallic lines are weaker and the double peaked profile is absent in the Balmer lines (Fig.~\ref{fig:ESPRESSO_LINES}). 
    According to the metallic lines and the veiling HM\,Lup has returned to a lower state of accretion in the epoch 4 ESPRESSO spectrum. The behavior of the metallic lines between the epoch 3 and 4 spectra (Fig.~\ref{fig:FeII_4924} and Sect.~\ref{line_variability}) suggests that the enhanced redshifted emission in the Balmer series at epoch 4 is caused by an increase in the emissivity in the outer region of the accretion flow, where these lines are produced. An alternative explanation could be a decrease in the opacity of this region, that has caused redshifted absorption at epoch 3.
    
    
	\section{Discussion}
	\label{discussion}
	
	\subsection{Temperature stratification of the accretion flow}
	\label{Tstrat}
	\begin{figure*}
		\centering
		\includegraphics[width=\linewidth]{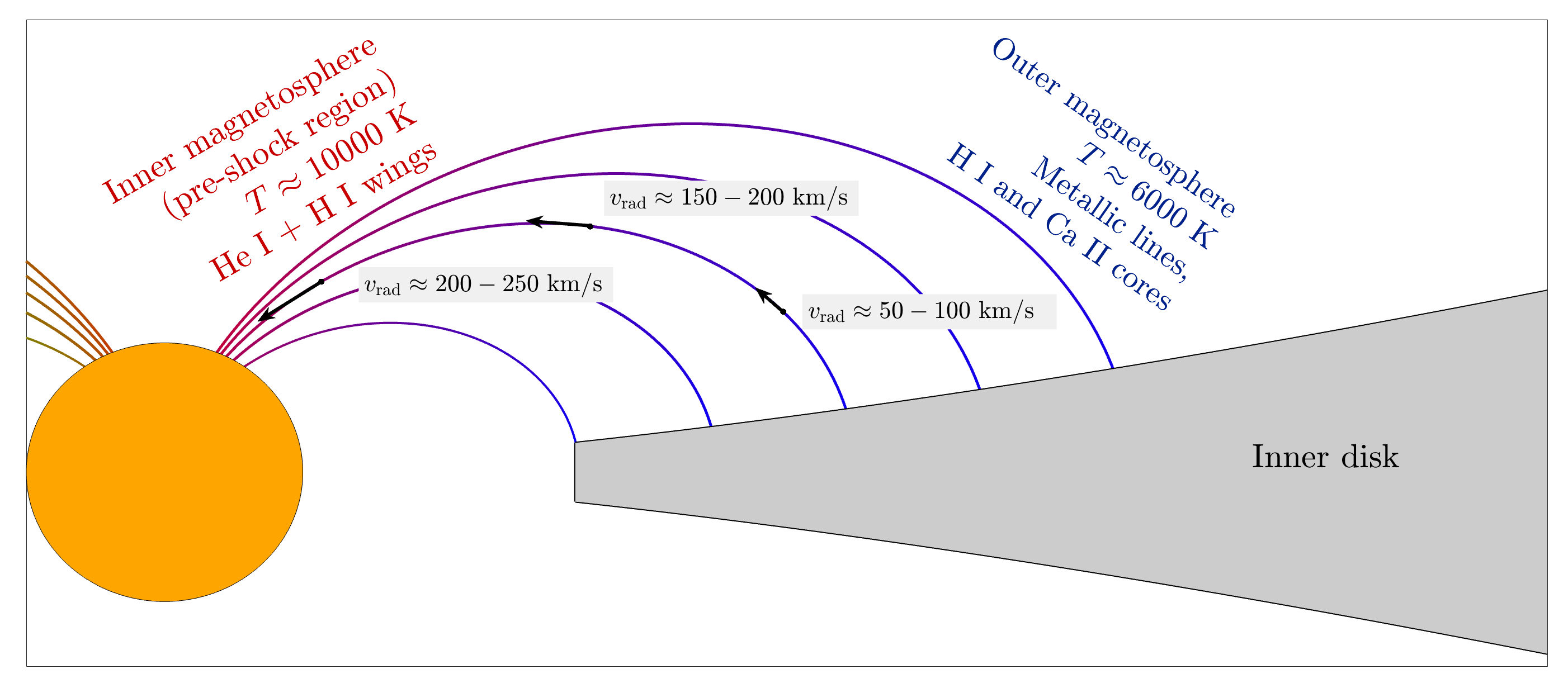}
		\caption{Temperature structure of the accretion flow that emerges from the line analysis. 
		}
		\label{fig:T_strat}
	\end{figure*}
	The many emission lines in the optical spectrum of HM\,Lup allowed us to probe the temperature stratification of the accretion flow. 
	Figure~\ref{fig:T_strat} shows a schematic representation of the structure that emerges from the line analysis. The \ion{He}{i} and \ion{H}{i} emission wings are produced in the pre-shock region, that we call the inner magnetosphere in Fig.~\ref{fig:T_strat}, in high temperature conditions ($T \approx 10000$~K). 
	The metallic lines originate in a more external region (near the disk) in which there is coexistence of Na, Mg, Ca, Ca$^+$, Fe, and Fe$^+$. This region, that we call the outer magnetosphere, has $T \approx 6000$~K.
	The strength of the \ion{Ca}{ii} lines relative to the \ion{H}{i} lines, as well as the observation of the \ion{Fe}{i} absorption against \ion{Ca}{ii}, suggests that the \ion{Ca}{ii} emission extends further inward, likely between the outer and inner magnetosphere.
	
	According to shock models, the pre-shock region is responsible for the Balmer continuum emission \citep{CalvetGullbring1998, Gullbring+2000}. Therefore, the \ion{He}{i} lines and the \ion{H}{i} line wings are related to the UV excess. 
	Since the \ion{Ca}{ii} lines are not produced in the pre-shock region, the fact that they overpredict the accretion luminosity might indicate a stronger emissivity from the outer part of the accretion flow in this system, relative to the stars in the \citet{Alcala+2017} sample.


	\subsection{Origin of the spectrophotometric variability}
    The analysis of the photometric data indicates that the magnetospheric interaction between the disk and the star is unsteady and highly dynamic in HM\,Lup. Three-dimensional magnetohydrodynamic (MHD) simulations showed that CTTSs may accrete in either a stable or an unstable regime \citep{Romanova+2003, Romanova+2004, KulkarniRomanova2008, Pantolmos+2020}. In the stable regime, accretion proceeds in two funnel streams and forms two polar hot spots on the stellar surface. In the unstable regime, the matter accretes in equatorial tongues and forms multiple hot spots on the surface of the star. The light curves are expected to be periodic with the stellar rotation period ($P_{\star}$) in the stable regime and stochastic in the unstable regime. 
    The transition between these two regimes depends on the ratio between the magnetospheric truncation radius $R_{\rm{T}}$ and the corotation radius $R_{\rm{co}}$ \citep{Blinova+2016}. Accretion is unstable if $R_{\rm{T}}/R_{\rm{co}} \lesssim 0.71$ and stable otherwise. 
    \citet{Blinova+2016} showed that when $R_{\rm{T}}/R_{\rm{co}}$
    decreases below $\sim 0.59$, unstable accretion becomes ordered. In the ordered regime, the matter accretes in one or two ordered tongues that rotate with the inner disk period.
    
    The comparison between the morphology of the two TESS light curves suggests that the accretion changed from a chaotic regime in 2019 to a more organized regime in 2021, which could be either the stable or the unstable ordered regime.
    The 2021 TESS light curve was in a quasi-periodic bursting state with $P = 4.79 \pm 0.28$~d. In the stable accretion scenario, this period would correspond to $P_{\star}$. The period detected in the third segment of the long term $B$ band AAVSO light curve, $P_{\rm{B3}} = 8.84$~d, is roughly twice the 2021 TESS period and can be interpreted as its first harmonic, that is easier to detect in the AAVSO light curve given the lower ($\sim 1$~d) cadence.
    However, using the $4.79$~d period as $P_{\star}$ and the $v\sin i$ and $R_{\star}$ values of Table~\ref{tab:stellar_accretion_pars} we obtain an inclination $i_{\star} = 23 \pm 6$~deg for the stellar rotation axis, that is, a star that is almost face-on. This is not in agreement with the observed quasi-periodic behavior of the light curve, that is likely the result of rotational modulation and therefore suggests a higher value for $i_{\star}$, as shown for instance by the 1D hydrodynamic simulations of \citet{Robinson+2021}.
    
    We propose that the quasi-periodic bursting behavior in the 2021 TESS light curve was the result of an increase in the accretion rate that compressed the magnetosphere and drove the system into a regime of unstable ordered accretion. 
    In this scenario, the $4.79$~d period detected in the LSP is not the stellar rotation period but it approximately corresponds to the timescale at which matter rotates at the truncation radius. At the boundary between the chaotic and ordered regime, the ratio of the stellar rotation period to the Keplerian rotation period at $R_{\rm{T}}$ is equal to $(R_{\rm{T}}/R_{\rm{co}})^{-3/2} = (0.45)^{-1} = 2.22$ ($\omega_s^{-1}$ in \citealt{Blinova+2016}). Therefore, $P_{\star}$ should be in the $8-10$~d range, compatible with $P_{\rm{B3}}$. Assuming this value as $P_{\star}$ we obtain $i_{\star} = 47 \pm 15$~deg, compatible with the observed quasi-periodic modulation. This value is also in agreement with the inclination of the outer disk axis (Table~\ref{tab:stellar_accretion_pars}) measured by \citet{Ansdell+2016}.
    The cycle-to-cycle variability of the bursts in the 2021 TESS light curve further supports this interpretation. According to the simulations by \citet{Blinova+2016}, the two tongues may carry different amounts of matter and one of them may sometimes disappear.
   
    The simultaneous spectroscopic data made it possible to characterize the photometric variability of the system during the main burst in the 2021 TESS observation in terms of variations in the accretion rate. 
    However, part of the line variability may be due to the rotational modulation of the accretion flow, as suggested by our analysis of the \ion{Fe}{ii}~4924 line variability and shown in simulations \citep[e.g.,][]{Kurosawa+2008, KurosawaRomanova2013}. Although the quasi-periodic behavior of the 2021 TESS light curve supports this scenario, our limited spectroscopic coverage does not allow to test the variability of the emission lines as a function of the rotational phase.
    We found that during the high accretion state the Balmer lines, except for H$\alpha$, develop an absorption component and the emission of low-ionization species increases. This is similar to what has been observed in VW\,Cha by \citet{Zsidi+2022b}. The dominant effect during the outburst is an increase in the density of the accretion flow. This produces a spectroscopic signature of the colder region near the disk, {that is}, the enhanced emission in the metallic species. 
    
	\section{Conclusions}
	We presented a comprehensive spectrophotometric study of the CTTS HM\,Lup. We examined the 2021 TESS light curve and the simultaneous spectroscopy obtained in the framework of the ULLYSES and PENELLOPE programs. The photometric data from 2021 were also compared with the 2019 TESS light curve and the long-term AAVSO monitoring. 
	
	The analysis shows that HM\,Lup is a star in which the accretion process is unsteady and rapidly variable. 
	Using different emission lines, we reconstructed the temperature structure of the magnetosphere.
    The \ion{He}{i} lines and the wings of the Balmer lines are formed in the pre-shock region and are related to the UV excess continuum. The emission lines from metallic species (\ion{Na}{i}, \ion{Ca}{i}, \ion{Ca}{ii}, \ion{Mg}{i}, \ion{Fe}{i}, and \ion{Fe}{ii}) are instead formed in lower temperature conditions, as already shown by \citet{Muzerolle+2001} for the \ion{Na}{i}~D lines.
    It must be stressed that the line forming region is more complex than our schematic picture. Density and temperature vary continuously along the accretion flow, and NLTE conditions are expected for the gas \citep{Azevedo+2006}. A detailed model should take into account the radiative transfer of light through a structure in which the local velocity and optical depth are functions of the distance from the star. The rich emission line spectrum of HM\,Lup makes this system ideal to test the temperature gradient in non-isothermal magnetospheric models.
	  
	The photometric behavior of HM\,Lup can be explained as the result of variations in the accretion rate in a system in which the magnetospheric radius is smaller than the corotation radius. 
	At the epoch sampled by the 2019 TESS light curve HM\,Lup was in an unstable and chaotic regime of accretion. During the TESS 2021 observation, the system entered into a regime of ordered accretion due to an increase in the accretion rate. The detected $4.79$~d period can be mistaken with the stellar period but it is likely associated to the Keplerian rotation of the matter at the truncation radius. We observed period variations in the long term AAVSO monitoring, in agreement with the simulations by \citet{Blinova+2016}. In particular, in the third segment of the $B$ band observations, the system has a periodicity of $8.84$~d, which could be the actual rotation period of the star.
	The spectrum of the system during the burst shows the following three main characteristics: (1) emission in the \ion{He}{i} lines increases without changes in the shape of the profile,
	(2) the Balmer lines develop an absorption component, (3) emission in the metallic species increases. 
	
	HM\,Lup gives the possibility to monitor a system with an outburst spectrum that is similar to, but less extreme than, that of EXors. 
	Additional observational material, such as an observing campaign capable of covering a timescale of $\sim 10$ days with high resolution spectra at nightly cadence and simultaneous photometry, is needed to better understand the nature of the complex variability of this Classical T Tauri Star.
	
	\bigskip

    \begin{acknowledgements} 
        This work has been supported by Deutsche Forschungsgemeinschaft (DFG) in the framework of the YTTHACA Project (469334657) under the project codes STE 1068/9-1 and MA 8447/1-1.
        AFR, and JMA acknowledge financial support from the project PRIN-INAF 2019 "Spectroscopically Tracing the Disk Dispersal Evolution" (STRADE) and the Large Grant INAF 2022 "YSOs Outflows, Disks and Accretion: towards a global framework for the evolution of planet forming systems" (YODA).
        CFM, JCW, and KM are funded by the European Union (ERC, WANDA, 101039452). Views and opinions expressed are however those of the author(s) only and do not necessarily reflect those of the European Union or the European Research Council Executive Agency. Neither the European Union nor the granting authority can be held responsible for them. 
        This work benefited from discussions with the ODYSSEUS team\footnote{\url{https://sites.bu.edu/odysseus/}} (HST AR-16129).
        The authors acknowledge Eleonora Fiorellino, Suzan Edwards, and René Oudmaijer for suggestions on this work. 
        AA acknowledges Steve Shore for valuable discussions on this system.
        Funding for the TESS mission is provided by NASA’s Science Mission directorate.
        The authors acknowledge with thanks the variable star observations from the {\it AAVSO International Database} contributed by observers worldwide and used in this research, and Elizabeth Waagen for coordinating the AAVSO Alerts.
        The authors acknowledge the use of the electronic bibliography maintained by the NASA/ADS\footnote{\url{https://ui.adsabs.harvard.edu}} system.
    \end{acknowledgements}

    \bigskip

	\bibliographystyle{aa} 
	\bibliography{HM_LUP.bib} 
	
	\begin{appendix}
	    \section{Flux calibration of the ESPRESSO spectra}
	    \label{ESPRESSO_flux_calib}
	    We flux-calibrated the four ESPRESSO spectra by re-scaling them to match the available $BVR_{\rm{c}}I_{\rm{c}}$ photometry. Since no photometric data were obtained simultaneously with the spectra, we linearly interpolated the AAVSO light curves to obtain the magnitude values at the ESPRESSO epochs. We treated the flux calibration as a least squares minimization problem, namely, we found the best scaling factor $r$ that matched the $BVR_{\rm{c}}I_{\rm{c}}$ photometry by minimizing the chi-square function
	    \begin{equation}
	        \chi^2(r) = \sum_{i = B,V,R_{\rm{c}},I_{\rm{c}}} \left[ \frac{m_i - M_i(r)}{dm_i} \right]^2
	        \label{chisquare}
	    \end{equation}
	    where $m_i$ and $dm_i$ are the observed magnitudes and uncertainties and $M_i(r)$ are the r-dependent "model" magnitudes, obtained by multiplying the original ESPRESSO spectra by $r$ and integrating over the passbands $i$ using the respective filter transmission curve.
	    The uncertainty on the scaling factor was estimated as the standard deviation of $\exp{(-\chi^2)}$.
	    Since the ESPRESSO spectra do not reach $9100$~{\AA}, the upper limit of the $I_c$ filter, we appended a rescaled version of the X-Shooter spectrum to the ESPRESSO ones long-ward of $7580$~{\AA}. The mean ratio of the two spectra between $7200$ and $7500$~{\AA} was used as scaling factor for X-Shooter. We repeated the fit without including the $I_c$ magnitudes, and the scaling parameter did not change within the uncertainties.
	    The obtained scaling factors are reported in Table~\ref{tab:ESPRESSO_scaling_factors}.
	    To check the reliability of this procedure, we compared the ESPRESSO spectrum at $\rm{MJD - MJD_0} = 3.83$~d to the X-Shooter spectrum, which was observed only $\sim 1$~h before. 
	    We computed the ratio of the two spectra, finding a mean flux ratio of $0.57$, in agreement with the scaling factor we obtained from the best fit procedure. 
	   
	    \begin{table}
    	\centering
    	\caption{Scaling factors for the flux calibration of the ESPRESSO spectra.} 
    	\begin{tabular}{lccc}
    		\hline
    		Spectrum & $\rm{MJD-MJD_0}$~[d] & $r$ & $\chi^2_{min}$ \\
    		\hline
    		ESPRESSO 1 & $2.89$  & $0.59 \pm 0.02$ & $64$ \\
    		ESPRESSO 2 & $3.83$  & $0.56 \pm 0.01$ & $2$ \\
    		ESPRESSO 3 & $5.75$  & $0.56 \pm 0.02$ & $25$ \\
    		ESPRESSO 4 & $12.86$ & $1.26 \pm 0.01$ & $13$ \\
    		\hline
    	\end{tabular}
    	\label{tab:ESPRESSO_scaling_factors} 
    \end{table}

    
    
    
    \section{Saha and Boltzmann equations}
    \label{LTE_eqs}
    In LTE, the population of the atomic levels of an ion is given by the Saha and Boltzmann equations \citep[e.g.,][]{HubenyMihalas2014}.
    The Saha equation provides the ratio of the populations of two successive ionization stages for a given atomic species
    \begin{equation}
        \frac{N_{\rm{j+1}}}{N_{\rm j}} = \frac{1}{n_{\rm{e}}} ~ \left(\frac{2\pi m_{\rm{e}} k_{\rm{B}} T }{h^2}\right)^{3/2} ~ \frac{2 U_{\rm{j+1}}(T)}{U_{\rm{j}}(T)} ~ e^{-\chi_j/k_{\rm{B}} T}
    \end{equation}
    where $T$ is the temperature, $N_{\rm{j}}$ and $U_{\rm{j}}$ are the number of atoms and the partition function of the $j^{\rm{th}}$ ionization stage, $n_{\rm{e}}$ is the electron density,  $m_{\rm{e}}$ is the electron mass, $h$ is the Planck constant, $k_{\rm{B}}$ is the Boltzmann constant and $\chi_j$ is the ionization potential of the lower ionization stage.
    The Boltzmann equation governs instead the population of the atomic levels in a given ion
    \begin{equation}
        \frac{n_{\rm{j}}}{N_{\rm{ion}}} = \frac{g_j}{U_{\rm{ion}}(T)} ~ e^{-E_j/k_{\rm{B}} T}
    \end{equation}
    where $n_{\rm{j}}$ is the population of the level with energy $E_j$ and multiplicity $g_j$ relative to the population of the ion $N_{\rm{ion}}$. 
    
    To compute the partition functions of the ions we obtained the energy levels and the ionization potentials from the NIST Atomic Spectra Database\footnote{\url{https://physics.nist.gov/PhysRefData/ASD/levels_form.html}}.
    If we assume that emission lines are optically thin, we can determine the properties of the emitting region by means of line ratios. The line strength can be defined as
    \begin{equation}
       j_{\rm{line}} = \frac{A_{\rm{ji}}}{\lambda_{\rm{line}}} ~ n_{\rm{j}}(n_{\rm{e}}, T)
    \end{equation}
    where $A_{\rm{ji}}$ and $\lambda_{\rm{line}}$ are the Einstein coefficient and the wavelength of the transition and $n_{\rm{j}}$ is the population of the upper level of the line, that depends on $N_{\rm{ion}}$ through the Boltzmann equation. The Saha equation gives $N_{\rm{ion}}$ as a percentage relative to the element population $N_{\rm{elem}}$.
    The line ratio between two lines from the same species is therefore a function of $n_{\rm{e}}$ and $T$. 
    
    In Sect.~\ref{metallic_lines}, we use this formalism to compute predicted LTE line flux ratios for a grid of $n_{\rm{e}}$ and $T$ values, with $10 \leq \log n_{\rm{e}} \leq 14$ and $4000~\rm{K} \leq T \leq 8000~\rm{K}$. The observed line ratios were statistically compared to the predicted LTE ratios by maximizing the likelihood function
    \begin{equation}
        \mathcal{L}(n_{\rm{e}}, T) = \exp\left[-\left(\frac{r_{\rm{o}} - r_{\rm{m}}(n_{\rm{e}},            T)}{dr_{\rm{o}}}\right)^2\right]
    \end{equation}
    where $r_{\rm{o}}$ and $dr_{\rm{o}}$ are the observed ratio and its uncertainty and $r_{\rm{m}}$ is the expected LTE ratio.

    \section{Triple Gaussian model}
    \label{FeII4924_fit}
    In Sects.\ref{FeI_fluorescence} and \ref{line_variability} we used a triple Gaussian model to fit the continuum-normalized emission lines in the spectrum of HM~Lup. The fit model is 
    \begin{equation}
        F_{v} = 1 + \sum_{i=1}^{3} C_{\rm{i}} ~ \exp{\left[\frac{(v-v_{\rm{i}})^2}{2\sigma_{\rm{i}}^2}\right]}.
    \end{equation}
    For each Gaussian $C_{\rm{i}}$ is the amplitude, $v_{\rm{i}}$ is the velocity center and $\sigma_{\rm{i}}$ is the standard deviation.
     
    \end{appendix}
	
\end{document}